\documentclass[a4paper,12pt]{article}
\pdfoutput=1
\usepackage{a4wide}
\usepackage[latin1]{inputenc}
\usepackage{dsfont}
\usepackage{amsfonts}
\usepackage{amsmath}
\usepackage{amssymb}
\usepackage{amsthm}
\usepackage{graphicx}
\usepackage[numbers, sort]{natbib}
\allowdisplaybreaks[1]
\usepackage{color}
\usepackage{ulem}
\usepackage{multirow}

\usepackage[small,bf]{caption}
\newcommand{\e}{\text{e}}
\newcommand{\be}{\begin{equation}}
\newcommand{\ee}{\end{equation}}
\newcommand{\beq}{\begin{equation}}
\newcommand{\eeq}{\end{equation}}
\newcommand{\bea}{\begin{eqnarray}}
\newcommand{\eea}{\end{eqnarray}}

\newcommand{\braket}[1]{\langle #1 \rangle}

\newcommand{\Vfull}{V_{\text{full}}}

\usepackage{commath}
\usepackage{mathcomp}
\usepackage{colortbl}
\usepackage{bbold}
\usepackage{commath}

\newcommand{\dn}{ {\scriptstyle \Delta} \mspace{-0.5mu} n}
\newcommand{\phio}{ \phi_{\scriptscriptstyle 0} }
\newcommand{\phic}{ \phi_c }

\newcommand{\Hubble}{ \mathcal{H} }

\DeclareMathOperator{\dt}{dt}
\DeclareMathOperator{\dphi}{d\phi}
\newcommand{\Vloop}{V_{\text{loop}}}

\newcommand{\emptyline}{ \vspace{15pt} }
\newcommand{\kahler}{K\"{a}hler }
\newcommand{\kahlerdriven}{K\"{a}hler-driven }
\usepackage{slashed}

\usepackage[below, above]{placeins}

\begin{document}

\begin{titlepage}

\vspace*{-15mm}
\vspace*{0.7cm}

\begin{center}

{\Large {\bf K\"{a}hler-driven Tribrid Inflation}}\\[8mm]

Stefan Antusch$^{\star\dagger}$\footnote{Email: \texttt{stefan.antusch@unibas.ch}} and
David Nolde$^{\star}$\footnote{Email: \texttt{david.nolde@unibas.ch}}

\end{center}

\vspace*{0.20cm}

\centerline{$^{\star}$ \it
Department of Physics, University of Basel,}
\centerline{\it
Klingelbergstr.\ 82, CH-4056 Basel, Switzerland}

\vspace*{0.4cm}

\centerline{$^{\dagger}$ \it
Max-Planck-Institut f\"ur Physik (Werner-Heisenberg-Institut),}
\centerline{\it
F\"ohringer Ring 6, D-80805 M\"unchen, Germany}

\vspace*{1.2cm}

\begin{abstract}
\noindent
We discuss a new class of tribrid inflation models in supergravity, where the shape of the inflaton potential is dominated by effects from the K\"{a}hler potential. Tribrid inflation is a variant of hybrid inflation which is particularly suited for connecting inflation with particle physics, since the inflaton can be a D-flat combination of charged fields from the matter sector. In models of tribrid inflation studied so far, the inflaton potential was dominated by either loop corrections or by mixing effects with the waterfall field (as in ``pseudosmooth'' tribrid inflation). Here we investigate the third possibility, namely that tribrid inflation is dominantly driven by effects from higher-dimensional operators of the K\"{a}hler potential. We specify for which superpotential parameters the new regime is realized and show how it can be experimentally distinguished from the other two (loop-driven and ``pseudosmooth'') regimes. 
\end{abstract}
\end{titlepage}

\tableofcontents
\newpage

\section{Introduction}

Inflation is a very successful paradigm for solving the horizon and flatness problems and for creating the almost scale-invariant, adiabatic perturbations that have been observed in the cosmic microwave background (CMB) 
\cite{Guth:1980zm,inf2,inf3,inf4}. However, it is not clear how cosmic inflation is realized within particle theory. Most importantly, the identity of the inflaton field, which drives the rapid expansion, is still unknown. Various models of inflation have been proposed in the literature, however often the inflaton is just an additional gauge singlet, rather disconnected from the rest of the theory.\footnote{Some notable exceptions are, for example, inflection point inflation in the Minimal Supersymmetric Standard Model \cite{Allahverdi:2006iq,mssmInflation}, Standard Model Higgs inflation \cite{smHiggsInflation, smHiggsInflation3} and GUT Higgs inflation \cite{gutHiggsInflation,gutHiggsInflation2}.}

A proposed framework for connecting inflation with particle physics is supersymmetric hybrid inflation \cite{susyHybrid,Dvali:1994ms,Linde:1997sj,RehmanHybrid} in which inflation is ended by a particle physics phase transition. The energy scale of this phase transition is around the Grand Unification (GUT) scale: $\Lambda_{\text{inflation}} \sim M_{\text{GUT}} \sim 10^{16}$ GeV. This inspires hope that hybrid inflation may be connected to the spontaneous breaking of a GUT symmetry \cite{Dvali:1994ms}. However, the inflaton itself must be a singlet in such models, and the predictions for the CMB fluctuations -- which depend on the properties of the inflaton -- usually cannot be related to other observables.

This issue was improved by the development of supersymmetric tribrid inflation \cite{sneutrinoHybrid,Antusch:2008pn,tribridBasic,Antusch:2009vg}, which is a variant of hybrid inflation where the inflaton itself can be charged under the symmetries of the particle theory \cite{Antusch:2010va,Antusch:2011ei}. In particular, it can be a charged matter particle and might have observable effects in the low-energy theory. This could make it possible to determine the inflaton couplings both from particle physics constraints and from measurements of the CMB at the same time. 
In models of tribrid inflation studied so far, the inflaton potential was dominantly generated by either one-loop corrections from the Coleman-Weinberg potential, or by small vacuum expectation values of non-inflaton fields like the waterfall field. We refer to these regimes as loop-driven and (pseudo)smooth \cite{pseudosmooth} tribrid inflation. 

The goal of this paper is to complete the discussion of variants of tribrid inflation by analyzing the third possibility, namely the case that tribrid inflation is driven dominantly by effects from higher-dimensional operators of the \kahler potential. Based on a generalized superpotential for tribrid inflation we specify for which parameters the \kahlerdriven regime is realized, calculate the slow-roll predictions and show how it can be experimentally distinguished from the other two regimes by a measurement of the running of the spectral index. We also discuss how the new class of inflation models might be embedded into supersymmetric GUT and/or flavour models.

The rest of the paper is structured as follows: We first motivate and introduce tribrid inflation in section \ref{sec:generalizedTribrid}. In sections \ref{sec:scalarPotential} -- \ref{sec:loops} we discuss the \kahlerdriven regime and derive the predictions using a power series expansion of the \kahler potential. Section \ref{sec:summary} summarizes our findings and compares our results for \kahlerdriven with loop-driven and pseudosmooth tribrid inflation. There we also discuss how our results can be used for model building, using tribrid inflation as a framework for building explicit particle physics models of inflation.

\section{Introducing Tribrid Inflation}
\label{sec:generalizedTribrid}

\subsection{Supersymmetric Hybrid Inflation}

Before we introduce tribrid inflation, let us review the basic properties of supersymmetric hybrid inflation \cite{susyHybrid}, which in its simplest form is given by the superpotential
\begin{equation}
\label{eq:hybridW}
  W = \kappa \Phi \left( H^2 - M^2 \right),
\end{equation}
where $\Phi$ is the inflaton field, $H$ is the so-called waterfall field, $M$ is a mass scale and $\kappa$ is a dimensionless constant. $M$ and $\kappa$ can be chosen real without loss of generality by global phase redefinitions of the fields.

The resulting scalar potential has a nearly flat valley along the $\Phi$ direction. The flatness along this direction is lifted by loop corrections from the Coleman-Weinberg potential and by Planck-suppressed operators in the \kahler potential. This generates a gentle slope suitable for slow-roll inflation. Below some critical value $\Phi_c$, however, the waterfall field $H$ gets a tachyonic mass. This triggers a waterfall transition which usually ends inflation very quickly \cite{hybridWaterfall}.

Hybrid inflation is attractive for various reasons. It is a small-field model of inflation, which means that field values stay well below the Planck scale. This allows to study it in an effective field theory (EFT) framework and predictions can be derived  using only a finite number of operators with low mass dimension, while operators of higher mass dimension are safely suppressed by the Planck scale. In addition, when comparing the model's predictions with the measured CMB spectrum, the mass scale $\braket{H} = M$ turns out to be of the order of the Grand Unification scale: $\braket{H} \sim M_{\text{GUT}} \sim 10^{16}$ GeV. This inspires hope that models of hybrid inflation could be related to some GUT phase transition.

\subsection{Supersymmetric Tribrid Inflation}

An important restriction of the conventional supersymmetric hybrid inflation model discussed above is that $\Phi$ must be introduced as a singlet under all symmetries, except for a possible R-symmetry, because the hybrid superpotential in eq.~\eqref{eq:hybridW} requires a linear term $\kappa M^2 \Phi$. As a gauge singlet, the inflaton is somewhat disconnected from the main particle theory: particles in the matter sector are mostly disqualified as inflaton candidates, because they are usually not singlets. A notable exception is the right-handed sneutrino \cite{Murayama:1992ua,sneutrinoHybrid} which may play a role in neutrino mass generation while being a gauge singlet within, for instance, the MSSM or SU(5) GUTs. In Pati-Salam unified models or SO(10) GUTs, even the right-handed sneutrino is charged under the gauge group and cannot be the inflaton of conventional supersymmetric hybrid inflation.

For model-building purposes, it is desirable to allow non-singlets as inflaton fields and still retain the attractive features of hybrid inflation. This motivates the introduction of the tribrid superpotential:
\begin{equation}
\label{eq:generalizedTribridW}
  W = \kappa S \left( H^l - M^2 \right) + \lambda H^m \Phi^n,
\end{equation}
where we used natural units with $M_{\text{Pl}} = (8\pi G)^{-1/2} = 1,$ and $l$, $m$, $n \geq 2$.\footnote{$l=1$ is not compatible with a waterfall transition, $m=1$ generates an overly steep inflaton potential and $n=1$ tends to lead to high-scale SUSY breaking and requires that $\Phi$ is a singlet, which removes one of the main benefits of tribrid inflation.} The singlet $S$ is no longer the inflaton, but an auxiliary field which will be stabilized near zero both during and after inflation. It is only required to induce the term $|H^l - M^2|^2$ in the scalar potential. The inflaton $\Phi$ is coupled to $H$ by a non-renormalizable operator with coupling constant $\lambda$; this term will provide the $\Phi$-dependent mass term of the waterfall field. Note that $\Phi$ no longer appears linearly in the superpotential, so $\Phi$ can be charged under symmetries, including gauge symmetries.\footnote{It has been suggested that using gauge non-singlet inflaton directions is problematic: one might expect large radiative corrections to its mass due to Feynman diagrams involving gauge interactions \cite{gaugeEtaProblem}. However, during inflation the gauge non-singlet inflaton has a large vacuum expectation value which breaks the gauge symmetry and makes the corresponding gauge bosons very massive. It has been shown that this can suppress the potentially dangerous loop contributions and make them completely negligible in tribrid models \cite{Antusch:2010va}.}

In this paper, we will restrict ourselves to models of inflation which are at least qualitatively similar to models of conventional supersymmetric hybrid inflation. This means, we will demand that $\Phi$ is the inflaton of single-field slow-roll inflation, and that inflation is terminated by a waterfall in $H$ when $m_H^2$ becomes negative.\footnote{The superpotential of eq.~\eqref{eq:generalizedTribridW} also features inflationary trajectories where $S$ or $H$ can be the inflaton, or where inflation happens along non-trivial multi-field trajectories. Although these are interesting models of inflation in their own right, they are not the subject of this paper.}

In general, we should also include the effects of the \kahler potential $K$. We are working with small-field models below the Planck scale, so we can expand the \kahler potential in powers of the fields:
\begin{equation}
\label{eq:generalizedTribridK}
  K = |\Phi|^2 + |H|^2 + |S|^2  +  \sum\limits_{i+j+k \geq 2} \kappa_{ijk} |\Phi|^{2i} |H|^{2j} |S|^{2k},
\end{equation}
where the coefficients in front of the quadratic terms are fixed by the normalization of the fields. The \kahler potential is assumed to depend only on the modulus squared of the fields, which may be enforced by a suitable choice of symmetries.

Models of tribrid inflation, with $W$ and $K$ as specified above, can generate a small slope for the inflaton in three different ways:
\begin{enumerate}
 \item By loop corrections from the Coleman-Weinberg potential,
 \item from small, non-vanishing vacuum expectation values for $H$ and $S$ already during inflation,
 \item or by Planck-suppressed operators from the \kahler potential.
\end{enumerate}
In principle, mixed cases are possible, where all three sources contribute. However, it often happens that one of these effects dominates while the others can be neglected or added as small corrections. We call these limiting cases the loop-driven, the the (pseudo-)smooth and the K\"{a}hler-driven regime.

The loop-driven regime has been studied e.g.\ in \cite{valerieInflation} including some effects from the \kahler potential. It is similar to conventional hybrid inflation both in its dynamics and in its predictions. One generally finds that $r \lesssim 0.01$, $\alpha_s \simeq 0$ and $\braket{H} \sim M_{\text{GUT}} \sim 10^{16}$ GeV. $n_s$ can take any value in the range still allowed by the WMAP7 data \cite{WMAP7}, depending on the choice of the \kahler potential.

The pseudosmooth regime has been explored recently in \cite{pseudosmooth}, though only for specific \kahler potentials and for $l=m$. For these cases, the results are similar to the loop-driven regime, apart from $n_s$, for which $n_s \gtrsim 0.96$ is predicted.

In the remainder of this paper, we study the K\"{a}hler-driven regime, which has not been analyzed so far. In section \ref{sec:scalarPotential}, we identify the parts of the scalar potential which are relevant for tribrid inflation. Section \ref{sec:slowrollKahler} is devoted to calculating the slow-roll predictions from the inflaton potential, and in section \ref{sec:kahlerInflationConstraints} we derive constraints from the waterfall dynamics. Afterwards, we can estimate that the loop corrections are in fact small for a large fraction of the constrained parameter space.

\section{Scalar Potential for \kahlerdriven Tribrid Inflation}
\label{sec:scalarPotential}

\subsection{Identifying the Relevant Terms}

We now discuss under which conditions \kahlerdriven tribrid inflation, along the lines discussed above, can be realized. To this end, the following requirements will be imposed:
\begin{itemize}
 \item The slow-roll predictions for the CMB spectrum are determined by the inflaton potential. These predictions must be compatible with experimental exclusion limits.
 \item During inflation, $H$ and $S$ must have masses above the Hubble scale $\Hubble$; otherwise we do not have single-field inflation in $\Phi$.
 \item $H$ must have a tachyonic instability for small $\Phi$ for the waterfall to happen.
\end{itemize}
To study these requirements, we decompose the potential:
\begin{equation}
  \Vfull = V(\Phi)  +  \frac{1}{2} m_h^2(\Phi) \, h_R^2  +  \frac{1}{2}m_{h_I}^2(\Phi) \, h_I^2  +  m_S^2(\Phi) \, \lvert S \rvert^2  +  ... \:,
\end{equation}
where $H = { \frac{ 1 }{ \sqrt{2} } } ( h_R  +  \operatorname{i} h_I )$ and where $m^2_h$ and $m^2_{h_I}$ are the squared masses of the real and complex part of $H$. We note that in the considered class of models, $m^2_{h_I} \geq m^2_{h}$. The interaction terms of the inflaton field which are quadratic in $H$ or $S$ have been absorbed in an inflaton dependence of the masses: $m_h^2 = m_h^2(\Phi)$ and $m_S^2 = m_S^2(\Phi)$. The dots denote terms which have been dropped because they are irrelevant for our discussion.

For single-field inflation, $S$ and $H$ must be stabilized at 0 during inflation;\footnote{In this paper, we demand that $H=S=0$ is stable during inflation. $H \neq 0 \neq S$ corresponds to tribrid inflation in the (pseudo-)smooth regime \cite{pseudosmooth}.} their masses are required to be larger than the Hubble scale $\mathcal{H}$: $m_S, m_h > \mathcal{H}$ for $\Phi \gg \Phi_{c}$. For inflation to end via a waterfall phase transition, $m_h^2 < 0$ for $\Phi < \Phi_{c}$. $h_I$ can be ignored; it is stabilized during inflation due to $m^2_{h_I} > m^2_h$.

If these necessary conditions are satisfied, slow-roll inflation depends only on the single-field inflaton potential
\begin{equation}
\label{eq:Vinflaton}
V(\Phi)  = \Vfull(\Phi, H, S)|_{H=S=0},
\end{equation}
and on the critical field value $\Phi_{c}$ which can be calculated from the potential by solving $m_h^2(\Phi_{c}) = 0$ for $\Phi_{c}$. We can then use $V(\Phi)$ and the critical field value $\Phi_{c}$ to derive the slow-roll predictions.

\subsection{Inflaton Potential}
\label{sec:calcInflatonPotential}

The supergravity scalar potential for chiral superfields can be calculated from $W$ and $K$:
\begin{equation}
\label{eq:localSusyVF}
 V_F = e^K\left( D_i K^{i \overline{j}} D_j^*  - \lvert W \rvert^2  \right),
\end{equation}
where $K^{i \overline{j}}$ is the matrix inverse of the \kahler metric $K_{ \overline{i}j }$, and
\begin{align*}
 K_{ \overline{i}j } = \dmd{K}{2}{ {X^*_i} }{}{ {X_j} }{},\quad
 D_i = \dpd{W}{ {X_i} } + W \dpd{K}{ {X_i} }.
\end{align*}
Inflation is assumed to happen along a D-flat direction, so D-term contributions are ignored.

The kinetic terms are
\begin{equation}
\label{eq:localSusyKinetic}
  \mathcal{L}_{\text{kin}} = K_{\overline{i}j} ( \partial_\mu X_i )^\dagger (\partial^\mu X_j).
\end{equation}
These kinetic terms are not yet canonical due to Planck-suppressed operators in the \kahler potential. During inflation, we can achieve canonical kinetic terms by a field transformation (see appendix \ref{appendix:canonicalNormalization} for details). This canonical normalization has the net effect of inducing higher-dimensional operators in our potential. Its effect on the results turns out to be very small, but it is nevertheless included for completeness and because it does not add any new complications.

\emptyline

We can now determine the inflaton potential $V(\Phi)$ from eqs.~\eqref{eq:localSusyVF} and \eqref{eq:canonicalNormalizationTrafo} and dropping all terms proportional to $H$ or $S$. We obtain the inflaton potential:
\begin{align}
  \label{eq:inflatonPotential}
V(\phi) &= V_0 \left(  1 + a \, \phi^2  + b \, \phi^4   +  c \, \phi^6  \right) + O(\phi^8),
\end{align}
with
\begin{subequations}
\begin{align}
 V_0 &= \kappa^2 M^4, \\
 \label{eq:kahlerAK}
 a &= \frac{1}{2} \left( 1 - \kappa_{101} \right), \\
 \label{eq:kahlerBK}
 b &= \frac18 +  a^{2} - \frac{a}{2} +  \kappa_{200}\left( \frac14 - \frac{2a}{3} \right) - \frac{ \kappa_{201} }{4},  \\
 c &= \frac1{48} - \frac{b}{2} + \kappa_{200} \left(  \frac18 - \frac{4b}{3}  \right) + \frac{\kappa_{300}}{8} - \frac{\kappa_{301}}{8} + O(a), \\
 \phi &= \sqrt{2} \lvert \Phi \rvert.
\end{align}
\end{subequations}
We have introduced the canonically normalized real inflaton field $\phi$ so that we can later use the usual slow-roll equations for real scalar fields. Note that for \kahler potential coefficients $\kappa_{ijk} \lesssim O(1)$, we expect $a$, $b$, $c \lesssim O(1)$.

As a further simplification, we will only keep terms up to order $\phi^4$ in the inflaton potential to derive the predictions for inflation:
\begin{equation}
\label{eq:phi2phi4V}
  V  \simeq  V_0 \left(  1 + a \, \phi^2 + b \, \phi^4  \right).
\end{equation}
The effects of the higher-order operator $c \,\phi^6$ are estimated in appendix \ref{appendix:higherDim}, and we find that they are usually small. We therefore have a simple model of inflation with the independent parameters $(a, b, V_0, \phic)$. This model is simple enough to be solved using robust analytic approximations.

\begin{figure}
  \centering
  \includegraphics[width=15cm]{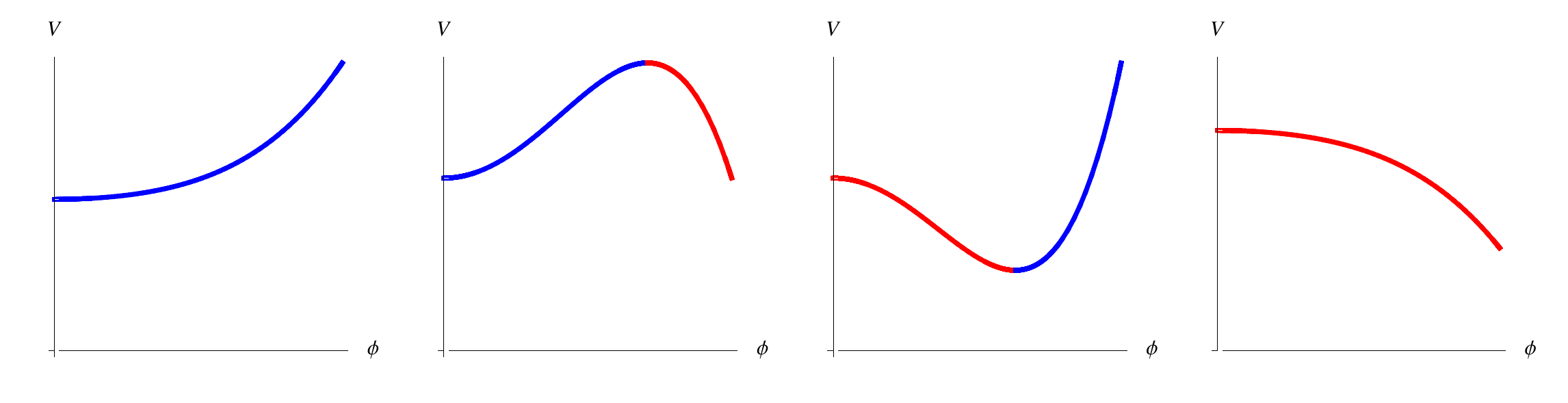}  
  \caption{Possible shapes of the approximate inflaton potential \eqref{eq:phi2phi4V}. The blue parts of the curve are suitable for inflation, whereas the red parts correspond to inverted trajectories which cannot be realized in supersymmetric tribrid models. For non-inverted trajectories, the hilltop-type potential (second from left) is the only possibility which is consistent with $n_s < 1$.}
  \label{fig:potentialShapes}
\end{figure}

The possible shapes of the potential are depicted in fig.~\ref{fig:potentialShapes}. The blue parts of the curve correspond to possible trajectories for non-inverted tribrid inflation, and the red parts correspond to inverted trajectories where the inflaton is rolling towards larger field values. We will later see that for supersymmetric tribrid inflation, inverted trajectories are not compatible with a waterfall transition. We will therefore consider non-inverted models, although the slow-roll treatment in section \ref{sec:slowrollKahler} can easily be generalized to the inverted case. We will also see that at about $N_0$ e-folds before the end of inflation, i.e.\ at the time where perturbations with wavenumber $k_0 = 0.002 \; \text{Mpc}^{-1}$ leave the horizon, $V''(\phio) < 0$ has to hold to account for $n_s < 1$ as required by the WMAP observations. We are therefore mostly interested in the hilltop-type potential with $a > 0$ and $b < 0$.

\subsection{Inflaton-Dependent Masses $m_h^2$ and $m_S^2$}

$m_h^2(\phi)$ and $m_S^2(\phi)$ can be determined from eqs.~\eqref{eq:localSusyVF} and \eqref{eq:canonicalNormalizationTrafo} analogous to the discussion in section \ref{sec:calcInflatonPotential}. The main difference is that we now have to keep terms quadratic in $H$ and $S$.

The masses induced by the \kahler potential $K$ are, using $V_0 = \kappa^2 M^4$:
\begin{subequations}
\begin{align}
\label{eq:onlyKahlerMassH}
 m_h^2 &= V_0 \left\{  1 - \kappa_{011}  +  \frac12 \phi^2 \left(   1  - \kappa_{011} - \kappa_{101} - \kappa_{111} + 2\kappa_{011} \kappa_{101} + \kappa_{011} \kappa_{110}   \right) +O(\phi^4)   \right\}, \\
\label{eq:onlyKahlerMassS}
m_S^2 &= V_0 \left\{  - 4 \kappa_{002}  +  \phi^2 \left(   \frac12 - 2\kappa_{002} - \kappa_{101} - 2\kappa_{102} + 6 \kappa_{002} \kappa_{101} + \frac{ \kappa_{101}^2 }{2}  \right) +O(\phi^4)  \right\}.
\end{align}
\end{subequations}
The masses induced by the superpotential $W$ depend on the parameters $l, m, n$. For $l, m \geq 3$ no additional mass terms are generated. Otherwise, the masses pick up extra terms:\footnote{$l=1$ is not considered because it generates a linear term for $H$ in the potential, destabilizing the metastable vacuum at $H=0$. This contradicts our assumption that $H=0$ before the waterfall. $m=1$ is also not considered, because it generates an overly steep inflaton potential inconsistent with slow-roll inflation.}
\begin{subequations}
\begin{align}
\label{eq:extraMassl2}
 l = 2: \quad \Delta m_h^2 &= \pm 2 \kappa^2 M^2 \left( 1 + O(\phi^2) \right), \\
\label{eq:extraMassm2}
 m = 2: \quad \Delta m_h^2 &= 2^{2-n} \lambda^2 \phi^{2n} \left( 1 + O(\phi^2) \right).
 \end{align}
\end{subequations}
The singlet field $S$ is stabilized with $m_S > \mathcal{H}$ if $\kappa_{002} < -\frac{1}{12}$, so it can safely be neglected during inflation. $\kappa_{002}$ does not appear in $V(\phi)$ or $m_h^2(\phi)$, so choosing a suitable $\kappa_{002}$ has no effect other than stabilizing $S = 0$.

The waterfall field's mass requires a more careful treatment. For $l > 2$ and $m > 2$, $m_h^2(\phi)$ is generally of the form
\begin{equation}
  \label{eq:onlyKahlerMasses}
  m^2_h(\phi) = V_0 \left[ \chi_0 + \chi_1 \phi^2 + ... \right], \quad(\chi_i \lesssim O(1)).
\end{equation}
In addition, $l=2$ gives a tachyonic mass to the real part of the waterfall field, and $m=2$ provides a positive $\phi$-dependent mass. These contributions dominate over the \kahler potential induced terms if $M \ll 1$ (for the $l=2$ term) and $\lambda^2 \gg V_0$ (for the $m=2$ term).

\section{Slow-roll Predictions}
\label{sec:slowrollKahler}

In this section, we deduce the slow-roll predictions for $\phio$, $\phic$, $V_0$, $r$ and $\alpha_s$ as functions of $(a, b, n_s)$. We choose $n_s$ as an input parameter instead of a prediction because the model can account for any spectral index, and treating it as an input parameter makes it easier to find the phenomenologically allowed parameter ranges.

We are interested in $\phio$ mostly as an intermediate step to derive all other predictions.\footnote{$\phio$ is the value of the inflaton field $N_0 \cong 50 - 60$ e-folds before the end of inflation, where perturbations with wavenumber $k_0 = 0.002 \; \text{Mpc}^{-1}$ leave the horizon.}
$\phic$ and $V_0$ will be needed to constrain $\braket{H}$ and $\lambda$, and $r$ and $\alpha_s$ are important cosmological observables which can be compared to current and future experiments.

As a pleasant side effect, we find a couple of useful constraints. In particular, we show that the vacuum energy must dominate throughout inflation ($V(\phi) \simeq V_0$) and that only the hilltop-type potential is suitable for inflation ($a > 0$ and $b < 0$).

This section only uses the inflaton potential from eq.~\eqref{eq:phi2phi4V}, so the results should be generally valid for hybrid models of hilltop inflation \cite{hilltop1,hilltop2} as long as loop corrections and higher-dimensional operators can be neglected. In particular, we predict that all such models have a running of the spectral index $\alpha_s \gtrsim 0$, and that a measurement of $\alpha_s$ precisely fixes the coefficient $a$ in the inflaton potential.

\subsection{Calculation of $\phio$}

In the slow-roll approximation, $\phio$ can be determined from the spectral index:
  \begin{align}
    n_s &= 1 - 6 \varepsilon_0 + 2 \eta_0 \notag \\
  \Leftrightarrow ~ 0 &= \left( 1 - n_s \right) - 3 \left( \frac{ V'(\phio) }{ V(\phio) } \right)^2 + 2 \left( \frac{ V''(\phio) }{ V(\phio) } \right) \notag \\
  \Leftrightarrow ~ 0 &= 4a + \dn +  \left( 24b - 8a^2 + 2a \dn \right) \phio^2  +  \left( \dn\, a^2 + 2\dn\, b - 20ab \right) \phio^4 \notag \\
      & \quad +  \left( -24b^2 + 2\dn\, ab \right) \phio^6 + \dn\, b^2\phio^8,
  \end{align}
where we have defined $\dn := 1 - n_s$. In this equation, several terms can be neglected by using that $\phio < 1$ (for the EFT to be valid) and $| \dn | \ll 1$ (experimentally from WMAP). Slow-roll inflation usually also requires that the inflaton mass term must be small: $m_\phi^2 \ll \Hubble^2 \sim V_0$, which for our potential means $a \ll 1$. In this paper, we assume $|a| \lesssim 0.1$; we have checked numerically that no consistent slow-roll solutions exist for larger $a$. The simplified equation is
  \begin{equation}\label{eq:phi0eq}
  0  \simeq  4a + \dn  +  24b \, \phio^2  -24b^2 \, \phio^6.
  \end{equation}
We can perturbatively calculate the approximate solutions for \eqref{eq:phi0eq}. The only consistent small-field solution is
\begin{equation}\label{eq:phi0}
  \phio^2 \simeq -\frac{4a + \dn}{24b}.
\end{equation}
The other solutions are either large-field solutions ($\phio \gg 1$) or they violate the slow-roll conditions ($\eta \gg 1$).

\subsection{Slow-roll Predictions for the CMB Spectrum}
\label{sec:kahlerSlowroll}

Eq.~\eqref{eq:phi0} implies that
  \begin{equation}
  b \, \phio^4 \simeq -\frac{4a + \dn}{24} \phio^2 \ll 1,
  \end{equation}
where we again used that $\dn \ll 1$, $\lvert a \rvert \lesssim 0.1$ and $\phi < 1$. This proves that the false vacuum energy $V_0$ must be dominant during the last $N_0$ e-folds:\footnote{We use that $\phi \leq \phio$ for non-inverted trajectories. This approximation would be invalid for inverted trajectories because then $\phi \geq \phio$ during inflation. In that case only the relation $V(\phi) = O(1) \, V_0$ would be enforced by the slow-roll conditions.}
  \begin{equation}
  \label{eq:vacuumEnergyDominant}
 V(\phi) = V_0 \big( 1 + \underbrace{a \, \phi^2}_{\ll 1} + \underbrace{ b\, \phi^4}_{\ll 1} \big) \simeq V_0.
  \end{equation}

The potential slow-roll parameters \cite{slowrollSeries} turn out to be small:
\begin{align}
\label{eq:epsilonPivot}
 \varepsilon(\phio) &= \frac12 \left( \frac{V'(\phio)}{V(\phio)} \right)^2 \simeq  \underbrace{ \frac89 \phio^2  \vphantom{\left(\frac{\dn}{8} \right)^2} }_{\lesssim O( 10^{-1} )} \underbrace{ \left( a - \frac{\dn}{8} \right)^2 }_{\lesssim O( 10^{-2} )}  \lesssim O( 10^{-3} ) \\
\label{eq:etaPivot}
 \eta(\phio) &= \left( \frac{V''(\phio)}{V(\phio)} \right) \simeq 2a + 12b\, \phio^2 \simeq  -\frac{\dn}{2} \simeq O(10^{-2}),\\
\label{eq:xiPivot}
 \xi(\phio) &= \left( \frac{V'(\phio) \, V'''(\phio)}{V(\phio)^2} \right) \simeq -\frac{16}{3} \left( a + \frac{\dn}{4} \right) \left( a - \frac{\dn}{8} \right).
\end{align}
In appendix \ref{appendix:slowroll}, we also show that the slow-roll parameters are small for all inflaton field values during inflation, which is required for the consistency of the slow-roll approximation.

\emptyline

\begin{figure}[tb]
\begin{minipage}[t]{0.5\textwidth}
  \centering
  \includegraphics[width=\textwidth]{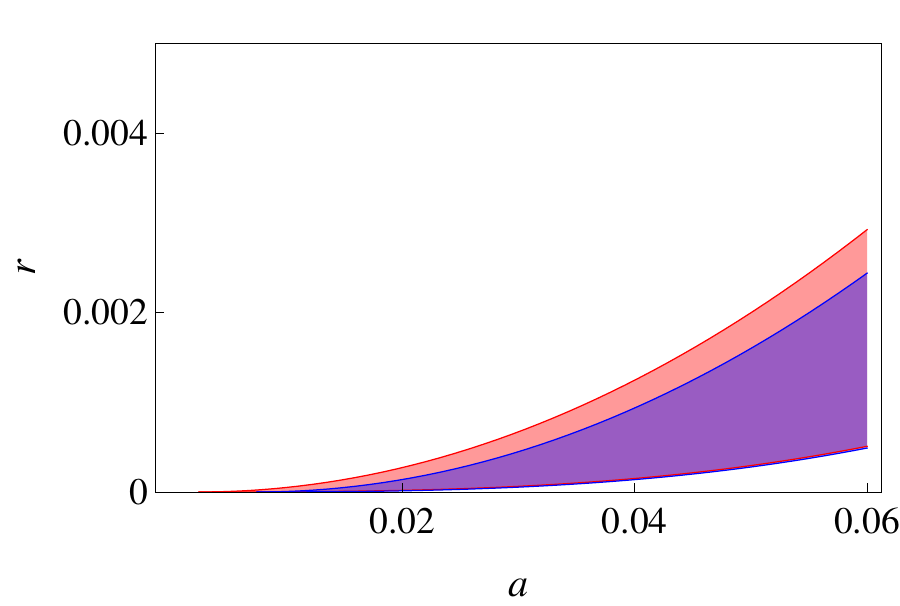}
\end{minipage}
\hfill
\begin{minipage}[t]{0.5\textwidth}
  \centering
  \includegraphics[width=\textwidth]{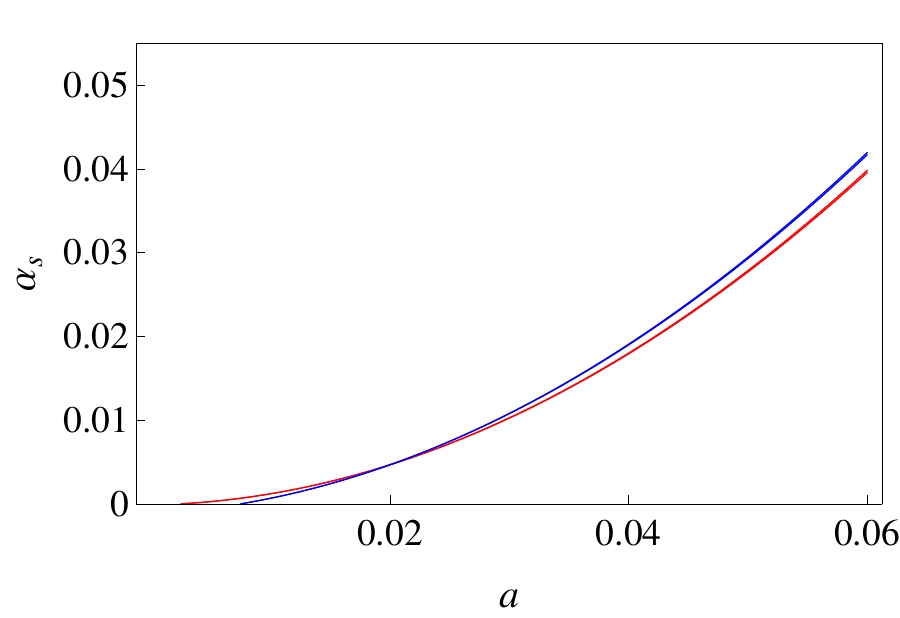}
\end{minipage}
  \caption{ \textit{Left:} Tensor-to-scalar ratio $r$. \textit{Right:} Running $\alpha_s$ of the spectral index. The blue and red bands correspond to $n_s = 0.94$ and $n_s = 0.98$; the upper and lower boundaries of the bands are given by $\phio < 0.25$ and $\lvert b \rvert < 1$. The tensor-to-scalar ratio is too small to be measured by the Planck experiment, and $\alpha_s$ is positive. As large positive $\alpha_s$ are disfavoured by WMAP \cite{WMAP7}, our result implies that $a \lesssim 0.04$ (95\% CL). Note that $\alpha_s$ is almost independent of $b$, so $a$ could be determined by a future measurement of $\alpha_s$.}
  \label{fig:kahlerRAlphaSResult}
\end{figure}

We can determine the predictions for the CMB spectrum using eqs.~\eqref{eq:epsilonPivot}~to~\eqref{eq:xiPivot}. The running of the spectral index is
\begin{align}
 \alpha_s &= \underbrace{ 16\, \varepsilon(\phio)\eta(\phio) }_{ \lesssim O( 10^{-4} ) } - \underbrace{ 24\, \varepsilon(\phio)^2 }_{ \lesssim O( 10^{-5} ) }  -  2\, \xi(\phio) \simeq  -2\, \xi(\phio)  \notag \\
 \label{eq:kahlerAlphaSResult}
 &\simeq  \frac{32}{3}\, a^2 - \underbrace{ \frac23 a \dn }_{ \lesssim O( 10^{-3} ) } + \underbrace{ \frac{\dn^2}{6} }_{ \lesssim O( 10^{-4} ) }  \simeq   \frac{32}{3}\, a^2  \, + \, O(10^{-3}).
\end{align}
For large $a > 0.04$, the running of the spectral index is positive and large, which is disfavoured by the WMAP data \cite{WMAP7}. This constrains $a$ to small values $a \lesssim 0.04$, where the running of the spectral index is within the present bounds.

The tensor-to-scalar ratio $r$ evaluates to
\begin{equation}\label{eq:kahlerTensorResult}
 r = 16\, \varepsilon(\phio) \simeq \frac{128}{9}\left( a - \frac{\dn}{8} \right)^2  \phio^2 \simeq 14.2 \, a^2 \, \phio^2 \, + \,  O(10^{-4}).
\end{equation}
Measurable tensor perturbations are not expected in most of the parameter space, only for large $a \simeq 0.1$ and $\phio \simeq 0.3$ we can get $r \simeq 0.01$. We have just excluded such large $a$ because they generate a large running $\alpha_s$ in contradiction with the experimental bounds. We therefore predict, as typical for hybrid and tribrid inflation, that \kahlerdriven tribrid inflation does not produce large tensor perturbations. If $r \gtrsim 0.01$ should be observed, this class of tribrid models would be ruled out (at least as predictive small-field models).

The numerical slow-roll results for $r$ and $\alpha_s$ are shown in fig.~\ref{fig:kahlerRAlphaSResult}. They agree well with our analytical estimate, including the quadratic dependence on $a$.

The final result is that $r$ is predicted not to be observed with the Planck satellite, and $\alpha_s$ is positive. If $\alpha_s$ was measured, this would fix the theory's free parameter $a$.

\subsection{Hilltop-Type Potential}
\label{sec:hilltopType}

So far, we have only assumed $\dn \ll 1$, which is required by the WMAP data independent of $r$ and $\alpha_s$. We have now found that $r$ is generally small, and $\alpha_s \gtrsim 0$. This implies $\dn = (1-n_s) > 0$ at over 95\% CL. From eq.~\eqref{eq:etaPivot} we also see that $\dn \simeq -2V''(\phio)/V_0$. We conclude that phenomenologically viable \kahlerdriven  tribrid inflation requires $V''(\phio) < 0$, which singles out the hilltop-type potential with $a > 0$, $b < 0$ as the only viable option out of all potential shapes in fig.~\ref{fig:potentialShapes}.

When inflation is realized with a hilltop-type potential, we have to make sure that we are on the blue part of the curve to have non-inverted inflation. This means we need to start inflation below the local maximum of the potential. The potential has its maximum at
\begin{equation*}
  V'(\phi_{\text{max}}) = V_0\left(  2a \, \phi_{\text{max}}  +  4b \, \phi_{\text{max}}^3  \right) \stackrel{!}{=} 0 
\end{equation*}
\begin{equation}
  \label{eq:hilltopMaximum}
  \Rightarrow~~ \phi_{\text{max}}^2 = -\frac{a}{2b}.
\end{equation}
$\phio$ must be below that value:
\begin{align}
&\phi_{\text{max}}^2 > \phio^2  \notag \\
\Rightarrow~~ & -\frac{a}{2b}  >  -\frac{4a + \dn}{24b} \notag \\
\label{eq:hilltopMaximumA}
\Rightarrow~~ & a  >  \frac{\dn}{8}.
\end{align}
We therefore found the constraints
\begin{equation}
a > \dn/8 > 0, \quad b < 0.
\end{equation}

\subsection{Vacuum Energy $V_0$ during Inflation}
\label{sec:V0kahler}

\begin{figure}[tb]
\begin{minipage}[t]{0.5\textwidth}
  \centering
  \includegraphics[width=\textwidth]{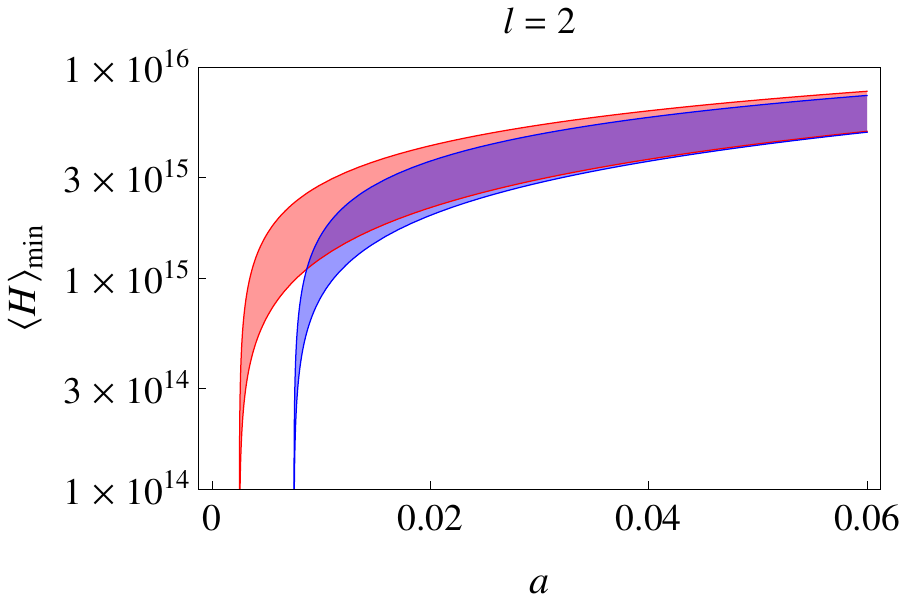}
\end{minipage}
\hfill
\begin{minipage}[t]{0.5\textwidth}
  \centering
  \includegraphics[width=\textwidth]{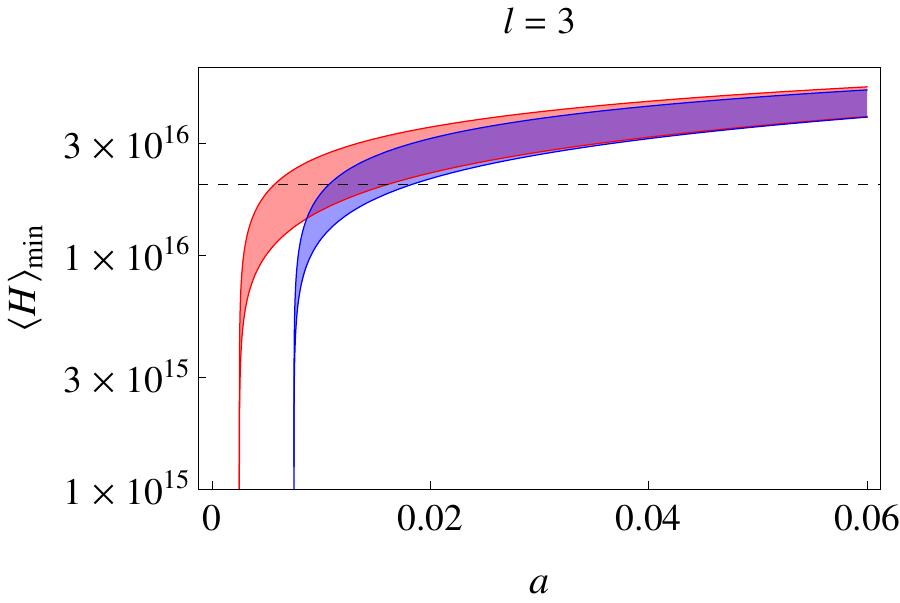}
\end{minipage}
  \caption{Lower bound on $\braket{H}$ after inflation for $l=2$ (left) and $l=3$ (right). The blue and red bands correspond to $n_s = 0.94$ and $n_s = 0.98$; the width of the bands is given by $\lvert b \rvert < 1$ and $\phio < 0.25$. Larger values of $\braket{H}$ are possible by choosing smaller $\kappa$, because $\braket{H} \propto \kappa^{-1/2}$; the plot shows the value for $\kappa = 1$. For $l > 3$, $\braket{H}$ can be determined from the plot for $l=3$ with eq.~\eqref{eq:kahlerHVevFromL3}. We see that in general, $\braket{H} \gtrsim O(M_{\text{GUT}})$, except for the spike at $a \simeq \dn/8$ where our approximations break down (see main text).}
  \label{fig:kahlerHiggsVev}
\end{figure}

The vacuum energy $V_0$ can be determined from the amplitude of scalar curvature perturbations $\Delta_{\mathcal{R}}^2$:
\begin{equation}
 \Delta_{\mathcal{R}}^2 = \frac1{24\pi^2} \left( \frac{V}{\varepsilon} \right)_{\phio} \simeq \frac1{24\pi^2} \left( \frac{V_0}{\varepsilon(\phio)} \right).
\end{equation}
If we replace the amplitude of curvature perturbations by its measured value $\Delta_{\mathcal{R}}^2  =  (2.43 \pm 0.09) \times 10^{-9}$ \cite{WMAP7}, we can solve for $V_0$ to find
\begin{align}
\label{eq:V0}
 V_0 &\simeq \frac{32\pi^2}{-9b} \Delta_{\mathcal{R}}^2 \left( a - \frac{\dn}{8} \right)^2 \left( a + \frac{\dn}{4} \right).
\end{align}
The prediction for $V_0$ can be translated into a lower bound on the vacuum expectation value $\braket{H}$ after inflation. The global minimum of the potential is at $H = M^{2/l}$, and $V_0 = \kappa^2 M^4$. For a given $V_0$, the condition $\kappa \lesssim O(1)$ provides a lower bound on $M$ and thus a lower bound on $\braket{H}$. This bound is shown in fig.~\ref{fig:kahlerHiggsVev} for $l=2$ and $l=3$. We see that $\braket{H} \gtrsim M_{\text{GUT}}$, except for the dip at $a \simeq \dn/8$ which we discuss below.

For larger values of $l$, $\braket{H}$ must be closer to $M_{Pl}$ because
\begin{equation}
\label{eq:kahlerHVevFromL3}
 \braket{H} = M^{2/l} = (M^{2/3})^{3/l} = (\braket{H}_{l=3})^{3/l}.
\end{equation}
For $l \geq 4$, $\braket{H} > M_{\text{GUT}}$ for most values of $a$.

\emptyline

We briefly want to discuss the dip at $a \simeq \dn/8$. We see from eq.~\eqref{eq:hilltopMaximumA} that this situation corresponds to $\phio \simeq \phi_{\text{max}}$; this means $\phio$ is almost exactly on the hilltop maximum. At the local maximum, $V'(\phi) = 0$ and therefore $\varepsilon = 0$. In the slow-roll approximation, $V_0 \propto \varepsilon(\phio)$, which explains why $V_0 \rightarrow 0$ for $a \rightarrow \dn/8$.

We should note, however, that in this region our approximations become less reliable. For $V'(\phi) \rightarrow 0$, even small contributions to $V'$ become important locally. Loop corrections from the Coleman-Weinberg potential may not be negligible in that region, and for very small $V'$ we also need to include e.g.\ soft SUSY breaking terms. In addition, $V_0 \rightarrow 0$ suppresses the whole tree-level inflaton potential, and therefore neglected terms of the full model's potential also become more important for the global shape of the potential. For $a \simeq \dn/8$, our approximations may therefore not be reliable, and we neglect this region from now on.\footnote{One might argue that while the relative error in $V'_{\text{exact}}/V'_{\text{approx}}$ may be large, this could be compensated by a slight shift in $\phio$ towards the maximum of the full model's effective potential, and that our results may still be approximately valid as long as the higher derivatives of the potential are still dominated by \kahler potential terms. However, our solution will certainly break down at very low scales when the potential becomes dominated by lower-energy physics, e.g.~soft SUSY breaking terms.}

\subsection{Critical Inflaton Value $\phic$}

The number of e-folds between $\phio$ and $\phic$ can be calculated by integration:
\begin{align}
  N_0 &= \int\limits_{t_{c}}^{t_0} \! \dt \mathcal{H} \simeq \int\limits_{\phic}^{\phio} \! \dphi \frac{V}{V'}  \simeq \int\limits_{\phic}^{\phio} \! \dphi \frac{V_0}{V'}  \notag \\
\label{eq:kahlerN0}
 &= \frac{1}{4a} \left[ \ln(\phi^2) - \ln \left\lvert a + 2 b \phi^2 \right\rvert \right]_{\phic}^{\phio}.
\end{align}
In the first line we have used that $V(\phi) \simeq V_0$, which has been proven earlier. Solving eq.~\eqref{eq:kahlerN0} for $\phic$, we find
\begin{align}
\phic^2 &\simeq \phio^2 \left( \frac{ a }{ e^{4a N_0}( a + 2b\phio^2 ) - 2b \phio^2 } \right) \notag \\
\label{eq:phic}
&\simeq \phio^2 \left(  \frac{ 12a }{ (8a - \dn) e^{4aN_0} + 4a + \dn }  \right).
\end{align}

We have now determined all the quantities we need to derive constraints on the tribrid model parameters, which we will do in the next section.

\section{Constraints on Model Parameters}
  \label{sec:kahlerInflationConstraints}

So far, we have derived constraints and predictions only from the inflaton field's dynamics during slow-roll inflation. We have not yet considered the dynamics of $H$.

We know that during inflation, $H$ must be stabilized, and that at the end of inflation, $H$ must develop a tachyonic mass to end inflation by a waterfall transition. We will use these conditions to constrain the superpotential parameters $l$, $m$ and $\lambda$. From this discussion it will also become clear why tribrid models cannot have inverted trajectories.

  \subsection{Constraint from the Waterfall Requirement}
  \label{sec:waterfallConstraints}

For inflation to end with a waterfall in $H$, we need a critical inflaton value $\phic$ with $m_h^2(\phic) = 0$. At the critical point, the $\phi$-dependent part of $m_h^2$ must exactly cancel the $\phi$-independent part. This constrains the tribrid superpotential to $m=2$.

We can understand this qualitatively\footnote{The condition $m=2$ can be proven more rigorously by explicit calculations, which we leave out for brevity.} from the form of eq.~\eqref{eq:onlyKahlerMasses}. A sufficiently fast waterfall transition requires that
\begin{equation}
\label{eq:MhGtrHubble}
  m_h^2(\phi) \gtrsim \Hubble^2 ~~ (\text{for } \phi \gg \phic), \quad m_h^2(\phi) \lesssim - \Hubble^2 ~~ (\text{for } \phi \ll \phic). 
\end{equation}
The $\phi$-dependent mass must therefore vary at least over ranges of $O(1)\Hubble$. Contributions from the \kahler potential induce only a small $\phi$-dependence of $m_h^2$:
\begin{equation}
\label{eq:onlyKahlerMassExcluded}
  \Delta m_h^2 \simeq V_0 \chi_1 \phi^2 \simeq \underbrace{3\chi_1}_{\lesssim O(1)} \underbrace{\phi^2}_{< O(1)} \Hubble^2 < \Hubble^2,
\end{equation}
which is too small for a waterfall transition. We therefore need a large $\phi$-dependent mass from the superpotential, which requires $m=1$ or $m=2$.

However, superpotentials with $m=1$ cannot provide a large mass because they induce large inflaton couplings, spoiling slow-roll inflation. We therefore find the necessary condition that $m=2$.

To show that $l \geq m = 2$ is also sufficient, we will evaluate $m_h^2(\phic) = 0$ for $m=2$ and different $l \geq 2$. This will also lead to constraints on $\lambda$.

\subsubsection{Inflaton-dependent Mass from the Superpotential: $l > 2, m = 2$}

\begin{figure}[htbp]
  \centering$
\begin{array}{cc}
\includegraphics[width=0.48\textwidth]{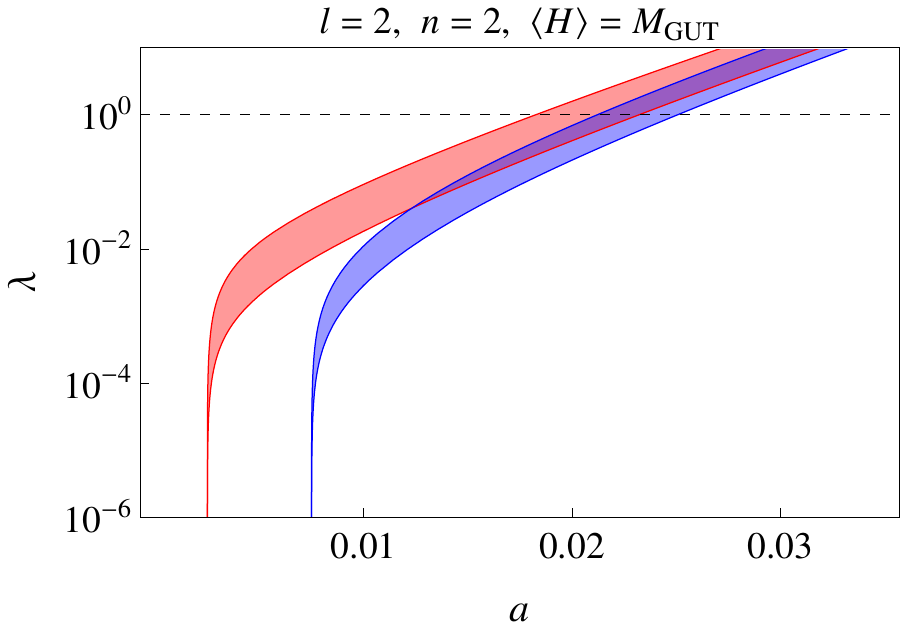} &
\includegraphics[width=0.48\textwidth]{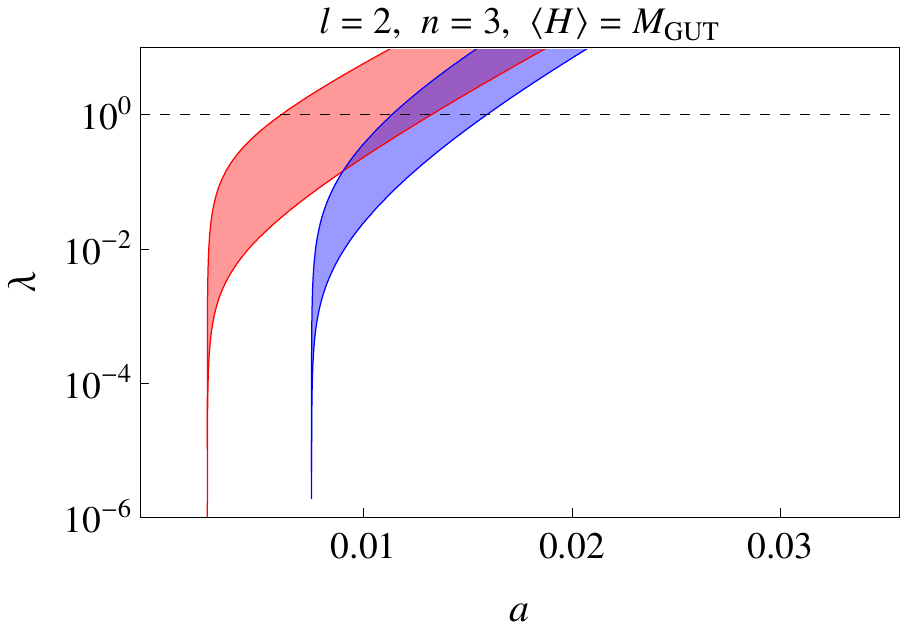} \\
\includegraphics[width=0.48\textwidth]{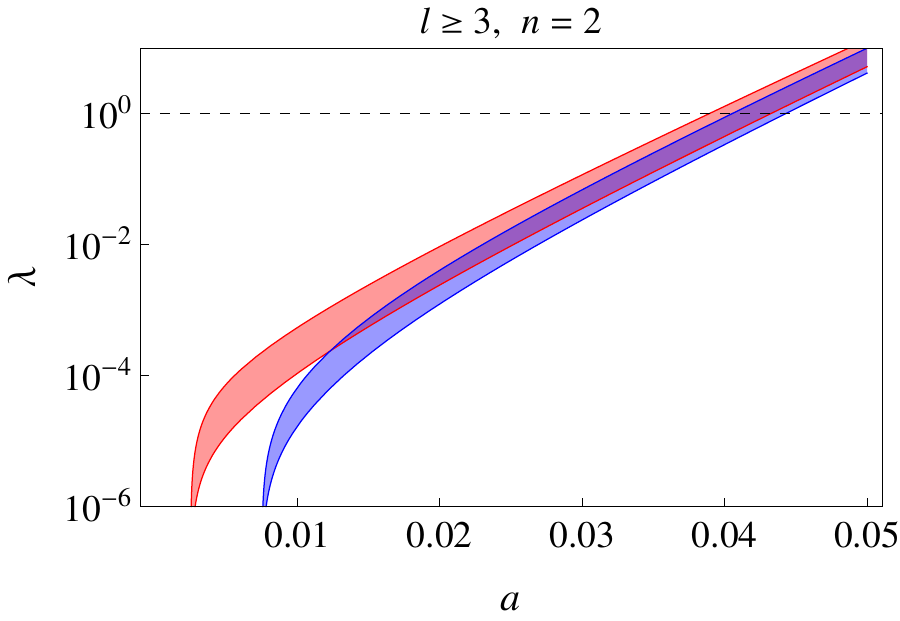} &
\includegraphics[width=0.48\textwidth]{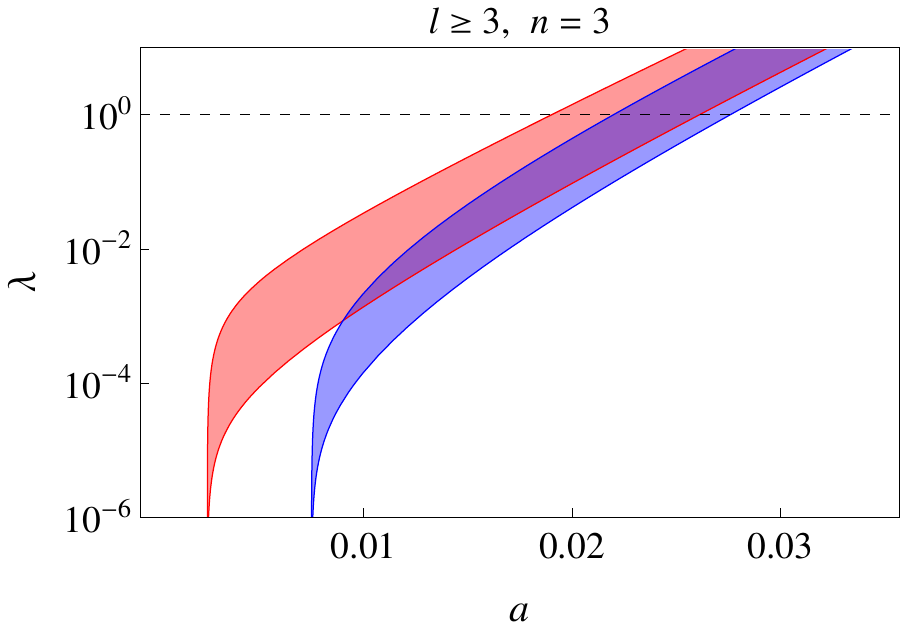}
\end{array}$
  \caption{Coupling constant $\lambda$ for various values of $l$ and $n$, assuming $N_0 = 55$. The blue and red bands correspond to $n_s = 0.94$ and $n_s = 0.98$; the width of the bands is given by $\lvert b \rvert < 1$ and $\phio < 0.25$. Note that for $l=2$, $\lambda \propto \braket{H}^{-1}$; in this case, the plot shows the value for $\braket{H} = M_{\text{GUT}}$. For $l \geq 3$, $\lambda \propto (\kappa_{011} - 1)^{1/2}$; for the plot we fixed $\kappa_{011} = 2$.}
  \label{fig:kahlerLambda}
\end{figure}

We insert eqs.~\eqref{eq:onlyKahlerMassH} and \eqref{eq:extraMassm2} in the waterfall condition $m_h^2(\phic) = 0$:
\begin{equation}
\label{eq:tempHMass1}
 \underbrace{2^{2-n}\lambda^2 \phic^{2n} + ...}_{\text{from superpotential}} + \underbrace{V_0 \chi_1 \phic^2 + ...}_{\text{from \kahler potential}}  \simeq  - V_0 \left(  1 - \kappa_{011}  \right).
\end{equation}
We know that the superpotential contribution must dominate on the left-hand side of eq.~\eqref{eq:tempHMass1}, because the \kahler potential contribution cannot provide a sufficiently large $\phi$-dependent mass. Eq.~\eqref{eq:tempHMass1} can therefore be approximated by
\begin{equation}
 2^{2-n}\lambda^2 \phic^{2n}  \simeq  V_0 \left(  \kappa_{011} - 1  \right).
\end{equation}
This fixes the coupling $\lambda$ to
\begin{equation}
  \label{eq:lambdaKahler3}
 \lambda^2  \simeq  2^{n-2} \left(  \kappa_{011} - 1  \right) \frac{ V_0 }{ \phic^{2n} }.
\end{equation}
We can now insert eqs.~\eqref{eq:V0} and \eqref{eq:phic} for $V_0$ and $\phic$ to determine $\lambda$. The possible values for $\lambda$ are shown in fig.~\ref{fig:kahlerLambda}. We find that the waterfall condition can be satisfied with a superpotential coupling $\lambda \lesssim 1$, while the tachyonic mass is generated from the \kahler potential. We note that this situation has appeared also in the first example of a tribrid inflation model in \cite{sneutrinoHybrid} (where $l=4$, $m=2$), however no detailed discussion of the inflationary dynamics and of the predictions was given.

The tachyonic mass $m_h^2(0)$ is of the order of the Hubble scale, which should be large enough to terminate inflation rather quickly. However, tighter constraints on $m_h$ could arise from the black hole bound \cite{hybridWaterfall}, which is hard to determine exactly but may be violated for small $m_h \lesssim \Hubble$.

\subsubsection{Conventional Tribrid Superpotential: $l = m = 2$}

The other case we consider is the usual tribrid superpotential with $l=m=2$. We insert eqs.~\eqref{eq:onlyKahlerMassH}, \eqref{eq:extraMassl2} and \eqref{eq:extraMassm2} in the waterfall condition $m_h^2(\phic) = 0$:
\begin{equation}
 \label{eq:tempHMass2}
 \underbrace{2^{2-n}\lambda^2 \phic^{2n} + ...}_{\text{from superpotential}} + \underbrace{V_0 \chi_1 \phic^2 + ...}_{\text{from \kahler potential}}  \simeq  \underbrace{ 2 \kappa^2 M^2 + ... \vphantom{V_0 \left(  1 - \kappa_{011}  \right)} }_{\text{from superpotential}} - \underbrace{ V_0 \left(  1 - \kappa_{011}  \right) }_{\text{from \kahler potential}}.
\end{equation}
On the right-hand side, the contribution from the superpotential is dominant because $M \ll 1$. On the left-hand side, we know it must be dominant because the $\phi$-dependent contribution from the \kahler potential is too small. We therefore find the condition:
\begin{equation}
 \label{eq:MlambdaCase1}
 2^{2-n}\lambda^2 \phic^{2n}  \simeq  2\kappa^2 M^2  =  2 \kappa \sqrt{V_0}.
\end{equation}
From eq.~\eqref{eq:MlambdaCase1} we can fix the coupling $\lambda$ as a function of $\kappa$:
\begin{align}
\label{eq:lambdaKahler2}
 \lambda^2 &\simeq 2^{n-1} \kappa \frac{ \sqrt{V_0} }{ \phic^{2n} }.
\end{align}
$\lambda$ is not fully determined by $a$, $b$, $\dn$ and $n$; it is also proportional to $\sqrt{\kappa}$. If we choose $\kappa$ such that $V_0 = \kappa^2 M^4 = \kappa^2 M_{\text{GUT}}^4$, we can find out what $\lambda$ we need for $\braket{H}=M_{\text{GUT}}$. The result is shown in fig.~\ref{fig:kahlerLambda}. Other choices of $\kappa$ lead to other values for $\lambda$ and $\braket{H}$, with $\lambda \propto \sqrt{\kappa} \propto \braket{H}^{-1}$.

\subsection{No Inverted Trajectories}

Based on the previous discussion, we can exclude inverted trajectories where $\phi < \phic$ rolls towards larger field values. For such inverted trajectories, the waterfall requires that $m_h^2(\phi)$ is positive for small $\phi$ and turns negative for large $\phi > \phic$. For this, the dominant $\phi$-dependent part must have a negative sign:
\begin{equation}
  \label{eq:proveNoInverted}
 m_h^2(\phi) = m_h^2(0) - \chi \, \phi^d + ... \quad(\text{for some $\chi$ and $d$}).
\end{equation}
We have seen that $m=2$ is required because during inflation the $\phi$-dependence of $m_h^2(\phi)$ must be dominated by the contribution from the superpotential operator $\lambda H^2 \Phi^n$, leading to a $\phi$-dependent mass term $\lvert 2\lambda H \Phi^n \rvert^2$. This mass term is strictly positive, so eq.~\eqref{eq:proveNoInverted} cannot be satisfied in non-(pseudo)smooth supergravity tribrid models.

\section{Smallness of Loop Corrections}
\label{sec:loops}

Throughout this paper, we have worked with the tree-level potential and ignored the one-loop corrections to the effective potential. We will now show that this is a good approximation for a large part of parameter space.

One-loop effects add the following extra term to the tree-level potential:
\begin{equation}
 \Vloop  = \frac{1}{64\pi^2} \sum\limits_{i} (-1)^{2s_i} m_i^4 \left[  \ln\left( \frac{m_i^2}{\mu^2} \right) - 3/2  \right],
\end{equation}
where $m_i$ is the $i$-th mass eigenstate, $s_i$ the spin of that eigenstate, $\mu$ the renormalization scale in the $\overline{MS}$ scheme and the sum runs over all physical degrees of freedom.

It is easy to show that $\Vloop \ll V$ due to the large constant term $V_0$ in the tree-level potential $V$. The derivatives are more constraining, because the large constant $V_0$ drops out when taking the derivative of $V$. We will therefore focus on the condition $\Vloop'(\phi) \ll V'(\phi)$. The magnitude of the loop corrections depends on the $\phi$-dependent masses. The dominant contribution stems from $m_h^2(\phi)$, as it receives a large $\phi$-dependent mass from the superpotential.

\begin{figure}[htbp]
  \centering$
\begin{array}{cc}
\includegraphics[width=0.48\textwidth]{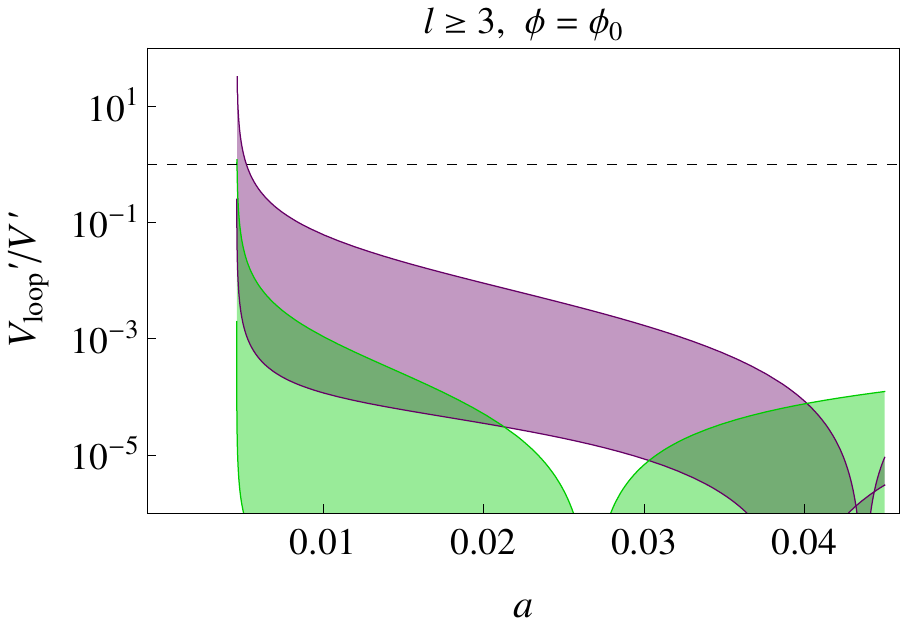} &
\includegraphics[width=0.48\textwidth]{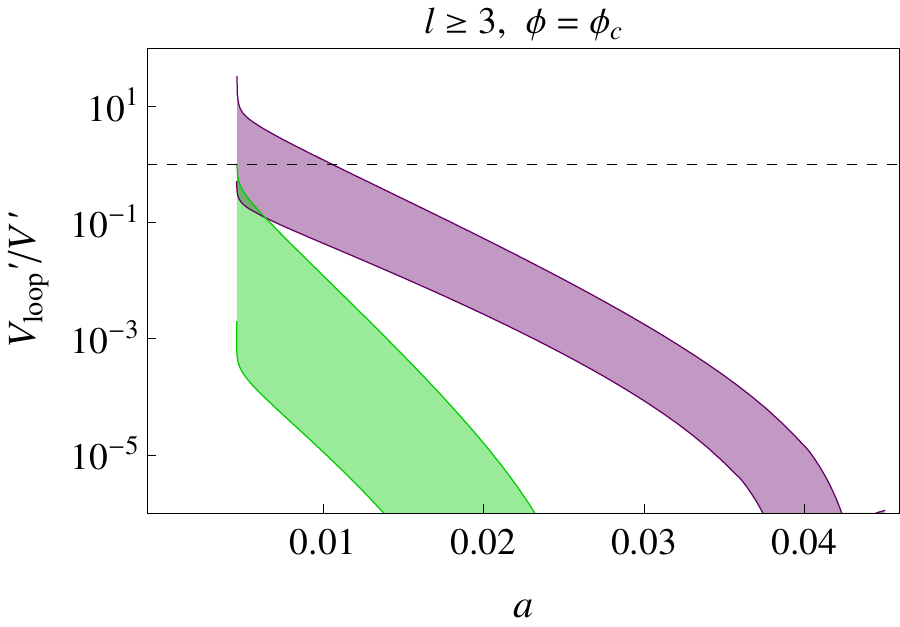} \\
\includegraphics[width=0.48\textwidth]{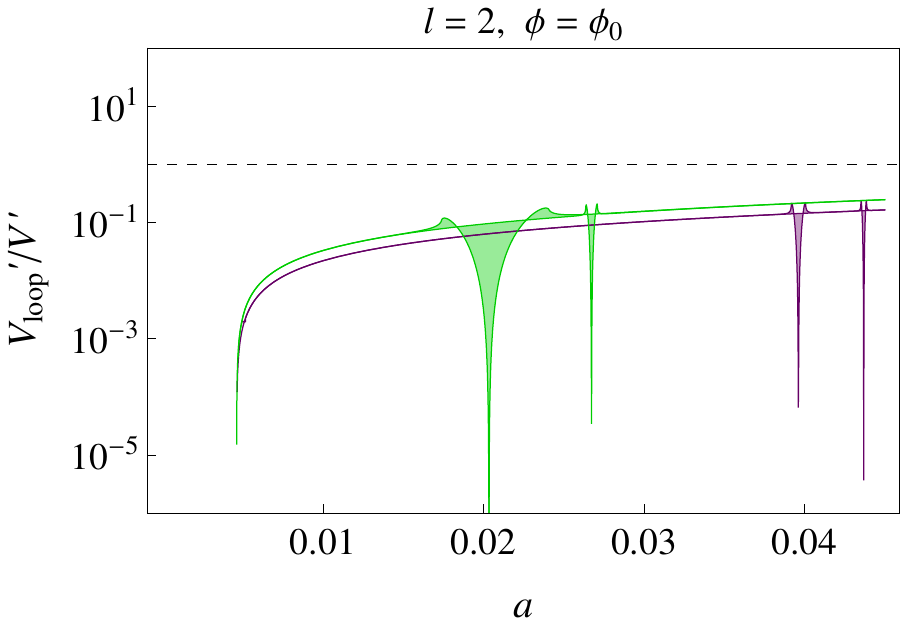} &
\includegraphics[width=0.48\textwidth]{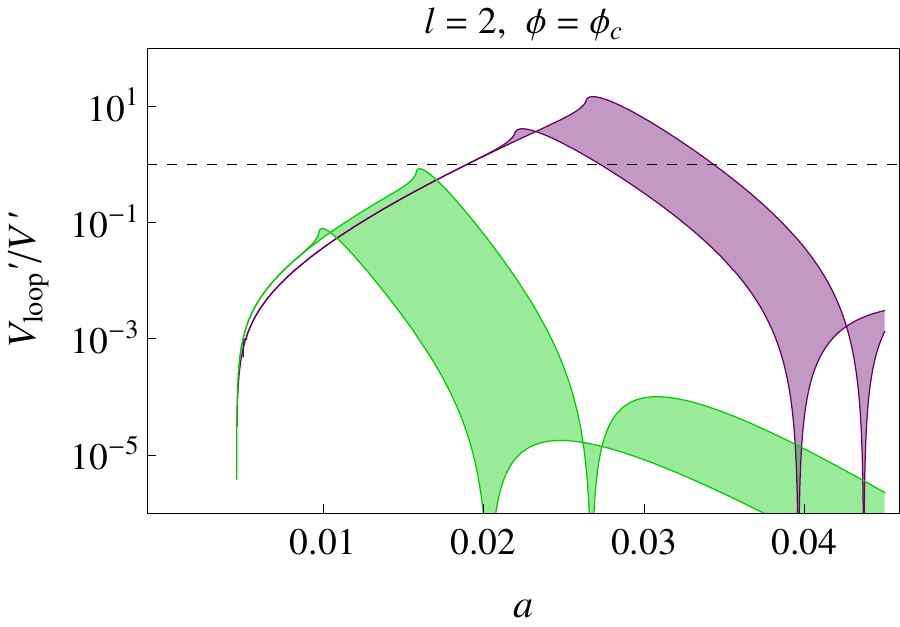}
\end{array}$
  \caption{Estimate of loop corrections $\Vloop'/V'$ for different values of $l$ and $\phi$ for $n_s = 0.963$, $N_0 = 55$ and $n=2$ (purple band) or $n=3$ (green band). The width of the bands is given by different choices of $b$: the lower boundary corresponds to $\lvert b \rvert = 1$ and the upper boundary to $b \ll 1$ (except for the small dips with $\Vloop' \rightarrow 0$, where the boundaries are sometimes reversed). For $l=3$, the plots estimate the loop corrections for generic \kahler potentials, whereas for $l=2$, specific \kahler potentials are required. Note that most predictions are only sensitive to the potential at $\phio$, where the loop effects are smaller; deviations near $\phic$ only change the prediction for $\lambda$. The lines in the lower plots show the extremal values of $b$; for continuous $b_{\text{min}} < \lvert b \rvert < 1$, the bands would fill the whole region between the spikes.\\
  We see that for $l = 3$, loop contributions are usually negligible, particularly for $b = O(1)$, $n = 3$ or $a \gtrsim 0.02$. For $l=2$, loop effects are often important. However, we find significant parts of parameter space, particularly for $b = O(1)$, $n = 3$ or $a \lesssim 0.01$, where loop effects can be neglected.}
  \label{fig:kahlerLoopPotential}
\end{figure}

\subsection{Loop Corrections for $l > 2$}

The tree-level masses $m_H$ and $m_F$ of $H$ and its fermionic superpartner have the form:
\begin{subequations}
\begin{align}
 x_H &= m_H^2(\phi) \simeq 2^{2-n} \lambda^2 \phi^{2n} + \chi_0 V_0 + \chi_1 V_0 \phi^2, \\
 x_F &= m_F^2(\phi) \simeq 2^{2-n} \lambda^2 \phi^{2n},
\end{align}
\end{subequations}
where $\chi_0$, $\chi_1 = O(1)$ parametrize the mass splitting between the scalar and fermionic components of $H$. This mass splitting is generated by Planck-suppressed operators from the \kahler potential. We ignore here canonical normalization effects and an overall factor of $\e^K$ as they induce only tiny corrections. They do not affect any cancellations because the scalar and fermionic components of the $H$ superfield get identical normalization factors.

We choose the renormalization scale $\mu^2 = 2^{2-n} \lambda^2 \phio^{2n}$ to make sure the logarithms are small during inflation, so that higher loop orders are properly suppressed. For representative values $\chi_0 = \chi_1 = -1$, we plot $\Vloop' / V_{\text{tree}}'$ for $\phi = \phio$ and $\phi = \phic$ (fig.~\ref{fig:kahlerLoopPotential}). Note that the lower bound of the band corresponds to $b = O(1)$, which may be considered a natural choice. We see that loop corrections are indeed small for most choices of $a$, with two exceptions: Loop effects may become important for $a \simeq \dn/8$ -- this case has been discussed in section~\ref{sec:V0kahler} -- and they can become important near $\phic$ if simultaneously $a \lesssim 0.02$ and either $n = 2$ or $b \ll 1$.

\subsection{Loop Corrections for $l = 2$}

For the case $m=l=2$, we recover the superpotential of conventional tribrid inflation. For this superpotential, it is known that the loop corrections can be significant. They are usually used to drive inflation in the absence of any tree-level potential for the inflaton. Nevertheless, as we will see, there is a complementary part of the parameter space where $\Vloop'$ is small and where \kahlerdriven inflation is realized.

If we neglect the SUGRA induced masses\footnote{The small mass splittings depending on the \kahler potential can have a significant impact on the loop corrections, because although they are orders of magnitude smaller than the superpotential-induced masses, they can make cancellations between scalar and fermionic contributions in the loop potential less effective \cite{valerieInflation}. To keep things simple, we concentrate here on a particular choice of \kahler potential where only the terms that we require for inflation are generated; this can be achieved by e.g.~choosing $\kappa_{011} = 1$ and $\kappa_{111} = 1 + \kappa_{110} - 2a$. For concrete models, one should plug in the whole \kahler potential and check whether the loop corrections are negligible for that particular model.}, we have
\begin{subequations}
\begin{align}
\label{eq:l2massesLoopEstimateScalar}
m_{H}^2 &= 2^{2-n} \lambda^2 \, \phi^{2n} \pm 2\kappa^2 M^2, \\
\label{eq:l2massesLoopEstimate}
  m_{F}^2 &= 2^{2-n} \lambda^2 \, \phi^{2n},
\end{align}
\end{subequations}
The ``$\pm$'' in eq.~\eqref{eq:l2massesLoopEstimateScalar} is ``$+$'' for the imaginary and ``$-$'' for the real scalar component of $H$. As we explained above, we can ignore canonical normalization effects and the factor $\e^K$ in both masses.

We can then plot $\frac{\Vloop'}{V_{\text{tree}}'}$ for $\phi = \phio$ and $\phi = \phic$ (see fig.~\ref{fig:kahlerLoopPotential}), using the same renormalization scale $\mu$ as above. We find that loop effects are often important, but there are still large parts of parameter space where they are suppressed, especially for $a \lesssim 0.01$ or $n = 3$. This shows that at least for some suitable \kahler potentials, loop effects can be negligible even with $l=m=2$.

\subsection{General Conclusion Concerning Loop Corrections}
For $l > 2$, loop corrections are usually negligible, particularly for $b = O(1)$, $n = 3$ or $a \gtrsim 0.02$. For $l=2$, loop effects are often important. We have nevertheless identified regions of parameter space where the tree-level description is sufficient,  especially if $b = O(1)$, $n = 3$ or $a \lesssim 0.01$. We conclude that for all $l \geq m = 2$, we find consistent models of \kahlerdriven tribrid inflation.

\section{Summary and Conclusions}
\label{sec:summary}

\subsubsection*{The three regimes of tribrid inflation}

The generalized tribrid superpotential of eq.~\eqref{eq:generalizedTribridW} can account for tribrid inflation in various ways. Apart from the formerly known regime where the inflaton potential is generated by loop corrections (see e.g.\ \cite{valerieInflation}) and the recently discussed pseudosmooth regime \cite{pseudosmooth}, we have shown that tribrid inflation can also happen in a \kahlerdriven regime where the inflaton potential is mostly generated by Planck-suppressed operators in the \kahler potential.

The three possible regimes, along with the required superpotential parameters $l$ and $m$, are shown in table~\ref{tab:summaryFinalML}. The K\"{a}hler- and loop-driven regimes require $l \geq m = 2$ to correctly end with a waterfall transition, whereas the pseudosmooth regime requires $l \geq m > 2$ to have suitable small-field trajectories.

\begin{table}[hbt]
\centering
  \begin{tabular}{ | l | c | c | c | c | }
    \hline
  & $m=1$ & $m=2$ & $m=3$ & $m = 4$ \\
    \hline
  \multirow{2}{*}{$l=2$} & \multirow{2}{*}{\textcolor{red}{not possible}} & \textcolor[rgb]{0,0.55,0}{K\"{a}hler-driven} & \multirow{2}{*}{\textcolor{red}{not possible}} & \multirow{2}{*}{\textcolor{red}{not possible}}   \\
  & & \textcolor[rgb]{0,0.55,0}{or loop-driven} & & \\
    \hline
  \multirow{2}{*}{$l=3$} & \multirow{2}{*}{\textcolor{red}{not possible}} & \textcolor[rgb]{0,0.55,0}{K\"{a}hler-driven} & \multirow{2}{*}{\textcolor[rgb]{0,0.55,0}{pseudosmooth}} & \multirow{2}{*}{\textcolor{red}{not possible}}   \\
  & & \textcolor[rgb]{1,0.5,0}{or loop-driven} & & \\
    \hline
  \multirow{2}{*}{$l=4$} & \multirow{2}{*}{\textcolor{red}{not possible}} & \textcolor[rgb]{0,0.55,0}{K\"{a}hler-driven} & \textcolor[rgb]{1,0.5,0}{pseudosmooth} & \multirow{2}{*}{\textcolor[rgb]{0,0.55,0}{pseudosmooth}}  \\
  & & \textcolor[rgb]{1,0.5,0}{or loop-driven} & \textcolor[rgb]{1,0.5,0}{or smooth} & \\
    \hline
  \end{tabular}
\caption{Viable (green) and dysfunctional (red) superpotential choices for supersymmetric tribrid inflation. The text indicates which regimes of tribrid inflation are possible for a given combination of $l$ and $m$. For the green entries, it has been explicitly shown that slow-roll inflation in agreement with observations is possible. The orange entries satisfy the necessary conditions and may or may not feature suitable slow-roll trajectories.}
\label{tab:summaryFinalML}
\end{table}

It is interesting to note that although all three regimes make similar predictions for $r \lesssim 0.01$ and $\braket{H} \gtrsim O(10^{16} \, \text{GeV})$, the \kahlerdriven regime can be distinguished from the other two regimes by a measurement of $\alpha_s$, which is predicted to be $\lvert \alpha_s \rvert < 10^{-3}$ \cite{pseudosmooth,valerieInflation} for the loop-driven and pseudosmooth regimes and $\alpha_s > 0$ for the \kahlerdriven regime.

\subsubsection*{\kahlerdriven tribrid inflation}

In this paper, the \kahlerdriven regime of tribrid inflation was analyzed for the first time in detail, using a generic expansion of the \kahler potential.

We noticed that the large number of model parameters can be mapped to an effective model with only three free parameters: $a$ and $b$, which can be calculated from the \kahler potential using eqs.~\eqref{eq:kahlerAK}--\eqref{eq:kahlerBK}, and the spectral index $n_s$, which will be measured with improved precision by the Planck satellite. Using these parameters, we analytically derived the slow-roll predictions for the tensor-to-scalar ratio $r$ and for the running of the spectral index $\alpha_s$ (fig.~\ref{fig:kahlerRAlphaSResult}). We also calculated constraints on the model parameters $\braket{H}$ (fig.~\ref{fig:kahlerHiggsVev}) and $\lambda$ (fig.~\ref{fig:kahlerLambda}) depending on $a$, $b$ and $n_s$ for the different choices of $l$ and $n$.

Successful inflation can occur for a large range of parameters. The required tuning of the \kahler potential, which is typical for supergravity theories, is pretty mild: it is sufficient to tune a single parameter $\kappa_{101} \simeq 0.94$--$0.99$, while all other parameters in the \kahler potential can take values of $O(1)$.

\emptyline

Our results can be used for model-building in various ways. If one knows the \kahler potential from some UV completion, one can calculate $a$ and $b$ using eqs.~\eqref{eq:kahlerAK}--\eqref{eq:kahlerBK}. The spectral index can be fixed by 
the upcoming measurement by the Planck satellite. One can then read off the constraints on the superpotential parameters $\lambda$ and $\braket{H}$ and the CMB observable $\alpha_s$ from our plots or calculate the numerical value from our analytical slow-roll results (eqs.~\eqref{eq:phi0}, \eqref{eq:kahlerAlphaSResult}, \eqref{eq:V0}, \eqref{eq:phic}, \eqref{eq:lambdaKahler3} and \eqref{eq:lambdaKahler2}).

If one does not know the \kahler potential, then $a$ could be fixed by a future measurement of $\alpha_s$. This suffices for an estimate of $\braket{H}$ and $\lambda$, as they do not strongly depend on $b$. One can also estimate the parameter $a$ by a measurement of the superpotential coupling $\lambda$; when tribrid inflation is embedded into a full particle physics model, this coupling may be observable in the low-energy theory. In sneutrino tribrid inflation, as discussed for example in the loop-driven regime (with $l = m = 2$) in \cite{valerieInflation}, $\lambda$ is related to the observable neutrino mass.

\emptyline

We took particular care to check our approximations: We explicitly showed that the slow-roll parameters remain small throughout inflation and that our semiclassical approximation is not spoiled by quantum fluctuations. We estimated the effects of higher-dimensional operators, which turn out to be small, and proved that loop effects are negligible for a large part of parameter space, especially for $l > 2$ (fig.~\ref{fig:kahlerLoopPotential}). We also compared our analytical approximations with numerical computations, and found excellent agreement. We therefore believe that our treatment covers the full range of K\"{a}hler-driven slow-roll tribrid inflation models with good accuracy.

\subsubsection*{Outlook}

Together with the papers on loop-driven (see e.g.\ \cite{valerieInflation}) and pseudosmooth \cite{pseudosmooth} tribrid inflation, this paper provides a blueprint for applying tribrid inflation to supersymmetric model-building. It has been shown that a variety of superpotentials can feature tribrid inflation, in particular all superpotentials of the form of eq.~\eqref{eq:generalizedTribridW} with $l \geq m = 2$ or $l=m>2$. It should now be possible to identify particle physics models which can contain the required superpotential terms. The inflaton can be a D-flat combination of (charged) matter fields \cite{Antusch:2010va}, and the waterfall phase transition might be identified with the spontaneous breaking of, e.g., a GUT symmetry \cite{Dvali:1994ms} or family symmetry \cite{Antusch:2008gw}.

Once a particle theory with a suitable superpotential has been found, the regime of tribrid inflation can be determined from $l$ and $m$ using table~\ref{tab:summaryFinalML}, and the inflationary predictions can be read off the graphs in the corresponding paper. As we have explained, this works particularly well for the K\"{a}hler-driven regime, where we have derived strong correlations between several model parameters and the cosmological observable $\alpha_s$.

We conclude that models of supersymmetric tribrid inflation are indeed promising candidates for realizing inflation in close contact with particle physics.

\section*{Acknowledgements}
This work was supported by the Swiss National Science Foundation.

\section*{Appendix}
\appendix

\section{Consistency of Assumptions}

Throughout this paper, we have treated our field in the semiclassical slow-roll approximation, using only the tree-level inflaton potential truncated after the $\phi^4$ term. In this appendix, we check the validity of these assumptions.

We will find that for a large part of parameter space, especially for $b = O(1)$, the approximations are very precise and our predictions are reliable. For another range of parameters, especially for $b \ll 1$, the quantitative results receive noticeable corrections, but the qualitative results are unchanged.

\subsection{Smallness of Slow-roll Parameters}
\label{appendix:slowroll}

To justify the slow-roll treatment, we must make sure that the slow-roll parameters $\varepsilon$ and $\eta$ are small for all values of $\phi \in [\phic, \phio]$. We start with $\varepsilon$:
\begin{equation}
\label{eq:smallnessSlowroll1}
 \varepsilon(\phi) = \frac12 \left( \frac{V'}{V} \right)^2 \simeq \frac12 \left( \frac{V'}{V_0} \right)^2 = \frac12 \left( 2a  \phi + 4b \phi^3 \right)^2,
\end{equation}
where we have used $V \simeq V_0$ from eq.~\eqref{eq:vacuumEnergyDominant}. We know that $a$, $\phi > 0$ and $b < 0$, so the two terms in the brackets partially cancel. We also know that $\phi^2 < \phi^2_{\text{max}} = -\frac{a}{2b}$ from eq.~\eqref{eq:hilltopMaximum}, so $\lvert 4b \phi^3 \rvert < \lvert 2a  \phi \rvert$. Together with eq.~\eqref{eq:smallnessSlowroll1} this implies the bound
\begin{equation}
\label{eq:smallnessSlowroll2}
 \varepsilon(\phi) < \frac12 \left( 2a  \phi \right)^2 = 2a^2 \phi^2 \ll 1.
\end{equation}
As $\phi < 1$ and $a < 0.1$, the first slow-roll parameter is small.

\emptyline

The second slow-roll parameter is
\begin{align}
 \lvert \eta(\phi) \rvert &= \left\lvert \frac{V''}{V} \right\rvert \simeq \left\lvert\frac{V''}{V_0} \right\rvert =  \lvert 2a + 12b \phi^2 \rvert  < \max \left\{ 2a, \lvert 12b \phi^2 \rvert \right\}  \notag \\
 &\leq \max \left\{ 2a, \lvert 12b \phio^2 \rvert \right\}  \simeq \frac{4a + \dn}{2} \ll 1,
\end{align}
where we used $\phi^2 \leq \phio^2 \simeq \frac{4a + \dn}{-24b}$. The second slow-roll parameter is also small due to $a < 0.1$.

The higher-order slow-roll parameters \cite{slowrollSeries} are also small due to the smallness of $V'$. We see that the slow-roll parameters are indeed always small, justifying our first-order slow-roll treatment.

  \subsection{Validity of Semiclassical Treatment}

\begin{figure}[htb]
  \centering
  \includegraphics[width=0.48\textwidth]{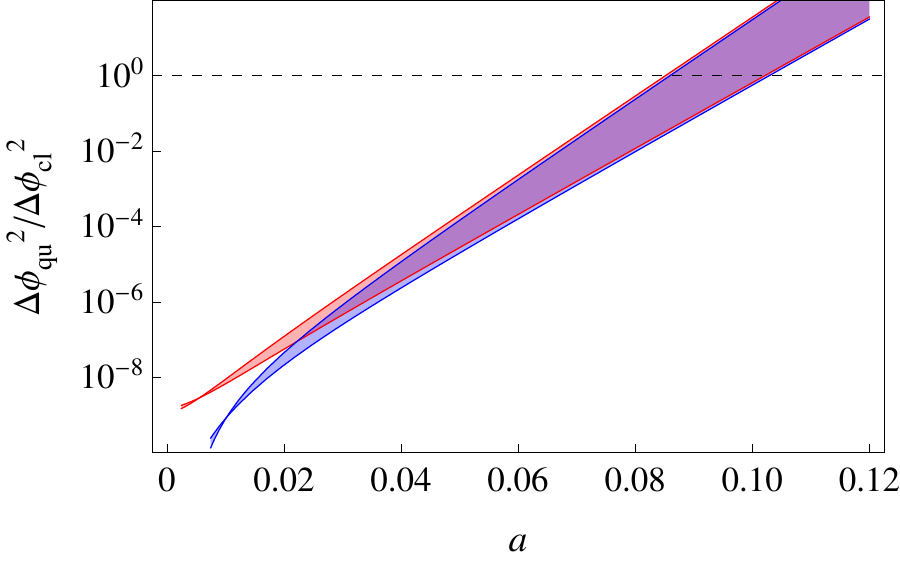}
  \caption{Ratio of quantum fluctuations $\Delta \phi_{\text{qu}}^2$ over the classical change of the field $\Delta \phi_{\text{cl}}^2$. The blue and red bands correspond to $n_s = 0.94$ and $n_s = 0.98$. The width of the bands is given by $N_0 = 50 - 60$. Note that we require $a \lesssim 0.04$ to keep $\alpha_s$ inside the WMAP $2\sigma$-bound. Therefore, quantum fluctuations are always negligible in the considered tribrid models, and our semiclassical approximation is justified.}
  \label{fig:semiclassicalValid}
\end{figure}

The semiclassical approximation assumes that quantum fluctuations can be neglected compared to the classical field dynamics:
\begin{align}\label{eq:quantumFluct}
 &\Delta \phi_{\text{cl}}  \gg  \Delta \phi_{\text{qu}} \notag \\
 \Leftrightarrow \quad& \varepsilon  \gg  \varepsilon(\phio) \Delta_{\mathcal{R}}^2.
\end{align}
For our potentials $\varepsilon$ is minimal at $\phio$ or $\phic$. At $\phio$, the condition \eqref{eq:quantumFluct} is trivially satisfied due to $\Delta_{\mathcal{R}}^2 = (2.43 \pm 0.09) \times 10^{-9}$, so we only need to check it at $\phic$. Therefore condition \eqref{eq:quantumFluct} simplifies to:
\begin{equation}
 \label{eq:semiclassicalCondition}
 \varepsilon(\phic)  \gg  \varepsilon(\phio) \Delta_{\mathcal{R}}^2.
\end{equation}
The numerical result for eq.~\eqref{eq:semiclassicalCondition} is plotted in fig.~\ref{fig:semiclassicalValid}. We see that for $a \lesssim 0.04$ -- which is required to keep $\alpha_s$ in the WMAP $2\sigma$-bound -- quantum fluctuations can safely be neglected.

\subsection{Effects of Higher-Dimensional Operators}
\label{appendix:higherDim}

\begin{figure}[htb]
\begin{minipage}[t]{0.48\textwidth}
  \centering
  \includegraphics[width=\textwidth]{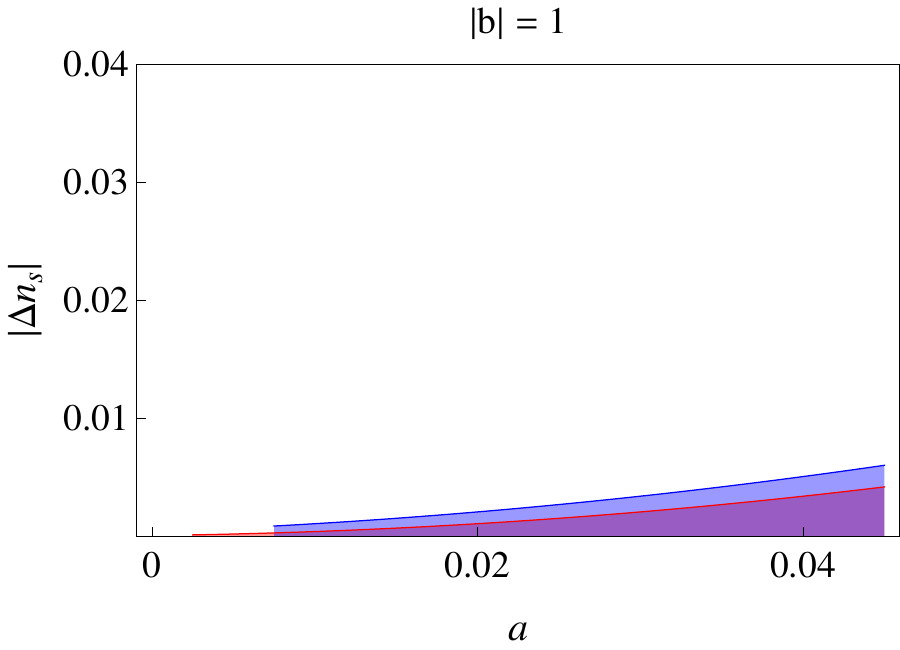}
\end{minipage}
\hfill
\begin{minipage}[t]{0.48\textwidth}
  \centering
  \includegraphics[width=\textwidth]{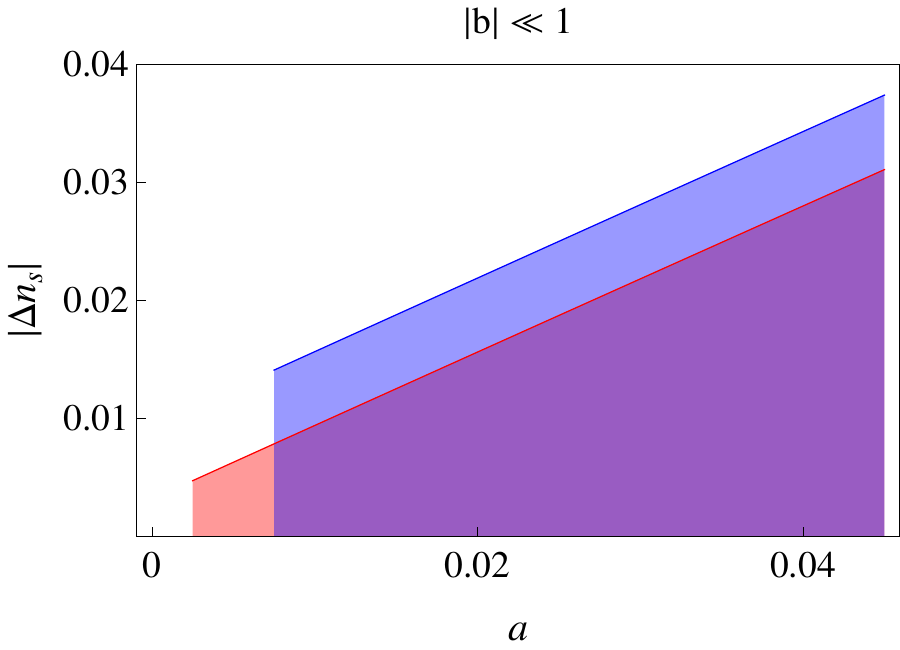}
\end{minipage}
  \caption{Upper bound on the shift of $n_s$ due to higher-dimensional operators for $\lvert b \rvert = 1$ (left) and $\lvert b \rvert \ll 1$ (right), assuming that $\lvert c \rvert \leq \lvert b \rvert$. The blue and red bands correspond to $n_s = 0.94$ and $n_s = 0.98$; the width of the bands is given by $\lvert c \rvert \leq \lvert b \rvert$ and $\phio < 0.25$. We see that for $b = O(1)$, the effect of higher-dimensional operators is negligible. For $b \ll 1$, the predictions can be shifted, but this shift is still smaller than the current experimental uncertainty in $n_s$.}
  \label{fig:kahlerHigherDimOperators}
\end{figure}

So far, we have neglected higher-order terms in the inflaton potential. We want to briefly discuss how these change the results of our analysis.

The most important neglected operator is $\Delta V = V_0 \, c \, \phi^6$, which is already strongly Planck-suppressed and does not contribute much to most of the results. Its strongest effect is on the higher derivatives of the potential:
\begin{equation}
 \Delta \eta \simeq \frac{\Delta V''}{V_0} \simeq 30c \, \phi^4.
\end{equation}
This shifts the spectral index:
\begin{equation}
 \Delta n_s \simeq 2 \Delta \eta(\phio) \simeq 60 c \, \phio^4.
\end{equation}
We can accommodate for such a change by using $\dn = n_s - 1 - \Delta n_s$ instead of our original definition $\dn = n_s - 1$ in all of our equations. However, for a large part of the parameter space, the effect turns out to be insignificant anyway. 

The maximal shift $\Delta n_s$ is shown in fig.~\ref{fig:kahlerHigherDimOperators}. For $b = O(1)$, the predictions are not changed significantly by higher-dimensional operators, and the results of this paper accurately describe \kahlerdriven tribrid inflation.

For $\lvert b \rvert \ll 1$, we assume that $\lvert c \rvert \lesssim \lvert b \rvert$. If $b$ is small due to some mechanism, it is likely that the same mechanism can suppress $c$. Under that assumption, we find that in the worst case the shift is as large as $\Delta n_s \sim 0.03$ (fig.~\ref{fig:kahlerHigherDimOperators}). This is still less than the separation between the red and blue bands in this paper's plots; it does not invalidate the qualitative results, but it blurs the quantitative predictions.

Only for $\lvert b \rvert \ll 1$ and $\lvert c \rvert > \lvert b \rvert$, the higher-dimensional operator $c \, \phi^6$ becomes dominant; results can change dramatically and $c$ must be included as a free parameter. In that case, the theory loses a lot of its predictivity. The qualitative result that $l \geq m = 2$ is still valid because it does not depend strongly on the inflaton potential, but the slow-roll predictions are invalidated.

\emptyline

One could also expect a large effect of $c$ on $\alpha_s$ via $\Delta V'''$, but this turns out to just change $\alpha_s \rightarrow \alpha_s(1 + 5 \frac{c}{b} \, \phio^2)$. Due to $\phio^2 \ll 1$, this does not affect the predictions too much for most values of $\frac{c}{b}$.

\section{Canonical Normalization Effects}
\label{appendix:canonicalNormalization}

This appendix deals with the effects of canonical normalization assuming that $H=S=0$ during inflation, as is the case during (non-pseudosmooth) tribrid inflation. In this case, we will find that canonical normalization has the net effect of inducing additional Planck-suppressed operators in the scalar potential.

\emptyline

\noindent In supergravity, the kinetic terms are given by
\begin{equation*}
  \mathcal{L}_{\text{kin}} = K_{\overline{i}j} ( \partial_\mu X_i )^\dagger (\partial^\mu X_j).
\end{equation*}
These kinetic terms are not yet canonical, though during inflation they are diagonal due to {$H=S=0$}.\footnote{In our case, the kinetic terms are diagonal if the \kahler potential contains no terms that are linear in $S$ or $H$, e.g.~$\Delta K = (S \lvert \phi \rvert^2 + h.c.)$ or $\Delta K = (HS + h.c.)$, which we have assumed in \eqref{eq:generalizedTribridK}. The off-diagonal elements involve a single derivative of $K$ with respect to $H$ or $S$. If $K$ contains no linear terms, the result must be proportional to at least one power of $H^\dagger$ or $S^\dagger$ (or their hermitian conjugates). Therefore, the off-diagonal elements vanish for $H=S=0$.} For $H$ and $S$, this can be solved by the simple transformations:
\begin{equation}\label{eq:kahlerTrafoHS}
 H = \frac{H'}{\sqrt{ \tilde{K}_{\overline{H} H } }}, \quad S = \frac{S'}{\sqrt{ \tilde{K}_{\overline{S} S } }}, \quad \tilde{K}_{\overline{i} j} \equiv \left( K_{\overline{i} j} \right)_{H=S=0}.
\end{equation}
The kinetic terms transform as
\begin{align*}
 \mathcal{L}_{\text{kin}}^{(H)} &= K_{\overline{H} H} \left\lvert \partial_\mu H \right\rvert^2\\
 &= K_{\overline{H} H} \left\lvert \frac{1}{\sqrt{ \tilde{K}_{\overline{H} H } }} \partial_\mu H' + H' \left( \dpd{}{\phi} \frac{1}{\sqrt{ \tilde{K}_{\overline{H} H } }} \right) \partial_\mu \phi  \right\rvert^2 \\
 &\xrightarrow{H=S=0}  \left\lvert  \partial_\mu H' \right\rvert^2,
\end{align*}
and analogously for $S'$. So we see that $H'$ and $S'$ are canonically normalized during inflation, where $H=S=H'=S'=0$.

Having normalized $H'$ and $S'$, we can next find a transformation $\Phi = f(\Phi')$ by demanding that $\Phi'$ should have canonical kinetic terms during inflation:
\begin{equation}
 \mathcal{L}_{\text{kin}}^{(\Phi)} = K_{ \overline{\Phi}\Phi }(\Phi) \left\lvert  \partial_\mu \Phi  \right\rvert^2 = K_{ \overline{\Phi}\Phi }( f(\Phi') ) \left\lvert  f'(\Phi') \partial_\mu \Phi'  \right\rvert^2  \stackrel{!}{=}  \left\lvert  \partial_\mu \Phi'  \right\rvert^2.
\end{equation}
For $H'=S'=0$, this condition reduces to an ordinary differential equation in $f$:
\begin{equation}
1 = \left\lvert \dod{f}{x} \right\rvert^2 \tilde{K}_{ \overline{\Phi}\Phi }( \Phi \mspace{-2.0mu} = \mspace{-2.0mu} f ).
\end{equation}
As our potential does not depend on the inflaton's phase, we can choose $\Phi$ to be real by a redefinition $\Phi \rightarrow \Phi \, \exp^{-i \operatorname{arg} (\Phi)}$. Then the differential equation can be solved up to any order by a power series ansatz for $f$. Using the \kahler potential \eqref{eq:generalizedTribridK}, the solution is
\begin{equation}\label{eq:kahlerTrafoPhi}
f(\Phi) = \Phi - \frac23 \kappa_{200} \, \Phi^3 + \left(  \frac{26}{15} \kappa_{200}^2 - \frac{9}{10} \kappa_{300}  \right) \Phi^5 + O( \Phi^7 ).
\end{equation}
After doing these transformations, we rename $\Phi'$, $H'$ and $S'$ back to $\Phi$, $H$ and $S$. The potential including normalization effects is then given by the potential \eqref{eq:localSusyVF} and the transformations \eqref{eq:kahlerTrafoHS} and \eqref{eq:kahlerTrafoPhi}:
\begin{equation}
 \label{eq:canonicalNormalizationTrafo}
  \Vfull = V_F \left( \Phi \rightarrow f(\Phi), H \rightarrow H / \sqrt{ \tilde{K}_{ \overline{H}H }( f(\Phi) ) }, S \rightarrow S / \sqrt{ \tilde{K}_{ \overline{S}S }( f(\Phi) ) } \right) .
\end{equation}

The effects of the field redefinition turn out to be negligible, as for $\kappa_{ijk} \lesssim O(1)$ it just transforms all operators in the potential
\begin{equation}
  \lvert \Phi^i H^j S^k \rvert  ~\rightarrow~  \lvert \Phi^i H^j S^k \rvert  \left( 1  +  O(1)\Phi^2 + O(1)\Phi^4 + ... \right).
\end{equation}
We see that the new operators arising from the transformation are strongly suppressed with respect to the already existing operators. We therefore expect that they do not affect the results significantly. They are nevertheless included in our analysis for completeness and because they do not add any extra complications.

\end{document}